\documentstyle[twocolumn,eqsecnum,aps,epsfig]{revtex}

\begin{document}
\draft
\title{Dynamical Chiral Symmetry Breaking on the Light Front\  I.\   
       DLCQ Approach} 

\author{K. Itakura}
\address{Research Center for Nuclear Physics, Osaka University, 
Osaka 567-0047, Japan,\\
 Yukawa Institute for Theoretical Physics, Kyoto University, 
 Kyoto 606-8502, Japan} 
\author{S. Maedan}
\address{Department of Physics, Tokyo National College of Technology, \\
Tokyo 193-8610, Japan}
\maketitle

\begin{abstract}
Dynamical chiral symmetry breaking in the DLCQ
  method is investigated in detail using a chiral Yukawa  model 
  closely related to the Nambu-Jona-Lasinio model.
By classically solving three constraints characteristic of the 
  light-front formalism, we show that
  the chiral transformation defined on the light front is equivalent 
  to the usual one when bare mass is absent. 
A quantum analysis demonstrates that a nonperturbative mean-field
  solution to the ``zero-mode constraint'' for a scalar boson 
  $\sigma$ can develop a nonzero condensate $\langle \sigma \rangle 
  =-\frac{\lambda}{N}\langle \bar \Psi\Psi \rangle \neq 0$ 
  while a perturbative solution cannot.
This description is due to our identification of the ``zero-mode constraint'' 
  with the gap equation. 
The mean-field calculation clarifies unusual chiral transformation properties 
  of fermionic field, which resolves a seemingly inconsistency between 
  triviality of the null-plane chiral charge $Q_5^{\rm LF}|0\rangle=0$ 
  and nonzero condensate $\langle \bar \Psi\Psi \rangle \neq 0$.
We also calculate masses of scalar and pseudoscalar bosons for both 
  symmetric and broken phases, and eventually derive the PCAC relation 
  and nonconservation of $Q_5^{\rm LF}$ in the broken phase.
\end{abstract}
\pacs{PACS number(s): 11.30.Rd, 11.30.Qc, 11.15.Pg, 12.40.-y}

\narrowtext

\section{Introduction}

Chiral symmetry breaking is undoubtedly one of the most important 
   concepts for understanding hadron physics in low energy region 
   \cite{Chiral}.
The smallness of $\pi$ and $K$ masses is beautifully explained 
   if one identifies them with the Nambu-Goldstone (NG) bosons 
   associated with the chiral symmetry breaking.
An important aspect of this phenomenon is dynamical formation 
   of NG bosons as bound states of quarks and gluons in the 
   strong coupling region.
However, its complete demonstration in QCD is not reached yet
   because of the difficulties in describing bound states in 
   a nonperturbative and relativistic manner.
Instead, many people have been 
   investigating much simpler effective models of QCD.
Among them, the Nambu-Jona-Lasinio (NJL) model \cite{NJL} is  
   the most deeply and thoroughly understood. 
The NJL model is a 3+1 dimensional four-Fermi theory and 
    reproduces various properties of hadrons concerning 
   the chiral symmetry breaking despite some undesirable features 
   such as nonrenormalizability and lack of confinement \cite{NJL_review}.
Nowadays the model plays a role of laboratory in which we can test
  new ideas proposed for nonperturbative study of low energy QCD. 
Therefore the NJL model is the most appropriate model in which we can 
  check whether the light-front (LF) quantization can be applied to the 
  dynamical chiral symmetry breaking.
The LF quantization is a newly revamped nonperturbative method
  for solving relativistic bound states in quantum field theory 
  \cite{Review}.

Let us explain why the chiral symmetry breaking 
  becomes a special issue in the LF formalism.
The reason is twofold: the first is apparent contradiction 
  between a nontrivial vacuum and a LF ``trivial'' vacuum,  
  and the second is peculiarity of LF chiral transformation.
To resolve these problems is our primary purpose in the present paper.
One of the remarkable merits of the LF quantization is that the Fock 
  vacuum defined by a free theory is also the vacuum of the full theory.
Many technical advantages such as exact Fock state expansion 
  arise from this fact.
On the contrary, the conventional formulation says that 
  the chiral symmetry breaking is essentially a physics of 
  finding another vacuum that breaks the chiral symmetry but is 
  energetically favored. 
Such ``vacuum physics'' is thought to be very important for 
  understanding nonperturbative phenomena in low energy region.
Therefore to apply the LF formalism to QCD necessarily entails 
  a problem how to realize such ``vacuum physics'' within a framework 
  with a trivial vacuum.
For the purpose of understanding this problem, there are considerable 
  efforts \cite{SSB} to describe the spontaneous symmetry breaking 
  in a simple scalar model ($\lambda \phi_{1+1}^4$). 
They succeeded in obtaining the critical coupling which is consistent 
  with the conventional results. 
The key is to solve a constraint equation for the longitudinal zero mode 
  (``zero-mode constraint'')  which appears 
  in the DLCQ (Discretized Light-Cone Quantization) method \cite{DLCQ}.
A nonzero condensate is realized as a 
  nonperturbative solution of the zero-mode constraint. 
We will discuss this method in more detail later.

Compared with such extensive studies,
   only little is known about the {\it dynamical} symmetry breaking 
   in {\it fermionic} systems.
Especially there have been only few attempts about the NJL model 
  on the LF \cite{LFNJL,Itakura,Heinzl_review,Bentz}.
At first glance, it seems not possible to follow the same route as in 
   the scalar models because we do not have bosonic fields as fundamental 
   degrees of freedom in the NJL model.
However, we can apply the same idea to the dynamical symmetry breaking 
   if one introduces {\it bosonic} auxiliary fields to the 
   fermion bilinears and raises them to dynamical variables by 
   adding their kinetic terms.
Of course the original fermionic model is reproduced as an infinitely 
   heavy mass limit of the bosonic fields.
According to this idea, we succeeded in describing the dynamical symmetry 
   breaking (discrete chiral symmetry) in the 1+1 dimensional 
   four-Fermi theory (the Gross-Neveu model) \cite{ItakuraMaedan}. 
The present paper is a generalization of this preliminary work which 
  discussed only {\it discrete} chiral symmetry.
We consider a kind of Yukawa model with {\it continuous} chiral symmetry, 
  which is obtained from the NJL model using the above technique.
We work within the DLCQ method so that {\it we can formulate the problem 
  from the viewpoint of the zero-mode constraints}.
It should be commented however that it is possible to discuss the dynamical
  symmetry breaking even without introducing auxiliary fields.
In Ref.~\cite{Itakura}, one of the authors insisted the importance of 
  a ``fermionic constraint'' which is again unique to
  the LF formulation and has very complicated structure 
  due to the four-Fermi interaction.
(Another merit of including scalar fields is a quite simplification 
  of the fermionic constraint.)
More detailed analysis in this direction will be reported in the next 
  paper \cite{II}.

One more point to be discussed is the unusual behavior of 
  chiral transformation on the LF.
In the LF formulation, a half degree of freedom of the fermion 
  is a dependent variable to be represented by other independent variables.
Therefore, chiral transformation should be imposed only on the 
  independent component of the fermion \cite{Mustaki}. 
It is not clear in interacting models whether the LF chiral 
  transformation is equivalent to the usual one or not.

The paper is organized as follows.
In the next section, we define the chiral Yukawa model which is closely 
   related to the NJL model and introduce our framework, the DLCQ method. 
The classical aspects of the model is discussed in Sec.~III. 
Here, we see that there are three constraints (i.e. two zero-mode 
  constraints and one fermionic constraint).   
We also show peculiarity of the LF chiral transformation and 
  explicitly give the null-plane chiral charge $Q_5^{\rm LF}$.
Quantum analysis, which is the main part of this paper, 
  is developed in Sections IV and V.
In Sec.~IV, we demonstrate that perturbative and nonperturbative treatments 
  of the solution to the constraints give different description of the model. 
In Sec.~V, we discuss some physics consequences of the nonperturbative 
  analysis.
Especially, we resolve a problem of contradiction between 
  the triviality of the null-plane charge and the nonzero condensate.
We further calculate the masses of scalars in the symmetric and broken 
  phases.  
Then we discuss the PCAC relations and nonconservation of the LF 
  chiral charge $Q_5^{\rm LF}$ in the broken phase.
The last section is devoted to conclusion and discussions.

\setcounter{equation}{0}

\section{The model}
Here we introduce the model (chiral Yukawa model) and summarize the standard 
  knowledge on the chiral symmetry breaking in the conventional 
  equal-time formulation.
We also define our setup of the problem following the DLCQ method.

\subsection{Definition of the model}
The NJL model was first introduced as the simplest $3+1$ dimensional 
  example which exhibits the dynamical chiral symmetry breaking 
  \cite{NJL}.
In its original form there were two flavors, but a one flavor model 
$$
 {\cal L_{\rm NJL}}=
   \bar \Psi^a ( i \rlap/ \partial - m )\Psi^a  
 + \frac{\lambda}{2N}
        \left[ 
              \left( \bar \Psi^a \Psi^a \right)^2
            + \left( \bar \Psi^a i \gamma_5 \Psi^a \right)^2  
        \right] 
$$
also breaks the chiral symmetry which exists 
 in the massless case $m=0$. 
We give an additional internal structure to the fermion
  independently of the flavor,
  and treat an $N$-component spinor $\Psi^a$ 
  in order to clarify the validity of approximation we use. 
 From now on, summation over $a=1,\ldots, N$ is always implied.
Since this model is not renormalizable,
  we must specify a regularization scheme such as a cutoff to uniquely 
  determine the model and to obtain finite results.

In this paper, we discuss more general model with 
  Yukawa interactions:
\begin{eqnarray}
  {\cal L} &=& \bar \Psi^a ( i \rlap/ \partial - m )\Psi^a 
             + {N \over 2 \mu^2} ( \partial_\mu \sigma
                   \partial^\mu \sigma + \partial_\mu \pi
                   \partial^\mu \pi )
                   \nonumber  \\
           & &  -{N \over 2 \lambda} (\sigma^2 +\pi^2)
                -(\sigma \bar \Psi^a \Psi^a 
                     + \pi \bar \Psi^a i \gamma_5 \Psi^a ), 
  \label{aa}
\end{eqnarray}
where  $\sigma$ ($\pi$) is a scalar (pseudoscalar) boson with mass 
  $\mu / \sqrt{\lambda} $ and $\mu$ is a dimensionless parameter.
If one takes infinitely heavy mass limit for scalars
   $ \mu \rightarrow \infty $, 
   the dynamical scalars become auxiliary fields 
   $\sigma=-\frac{\lambda}{N}\bar\Psi\Psi$,
   $\pi=-\frac{\lambda}{N}\bar\Psi i \gamma_5\Psi$ and the model 
   goes back to the NJL model.

In order to contrast with our LF calculation, 
 let us briefly comment on the usual story of 
 chiral symmetry breaking in the NJL model \cite{NJL}.
When $m=0$, both of the lagrangian densities are invariant 
 under the chiral transformation: 
\begin{eqnarray}
  \Psi^a   &\rightarrow&  {\rm e}^{ i \gamma_5 \theta } \Psi^a, 
  \label{ace}\\
%
%
   \left( \matrix{\sigma  \cr \pi} \right)
     & \rightarrow   &
   \left(
     \matrix{ \cos 2 \theta   &  \sin 2 \theta  \cr
             -\sin 2 \theta   &  \cos 2 \theta
	    }
   \right)
   \left( \matrix{\sigma  \cr \pi} \right)
  \equiv  R( 2 \theta ) 
   \left( \matrix{\sigma  \cr \pi} \right).
  \label{acf}
\end{eqnarray}
It should be reminded that this transformation is, of course, 
   imposed on all the fields, which is, however, not the case in 
   the LF formalism.
This point will be discussed later in more detail.
The usual story is as follows:
The chiral symmetry breaks down spontaneously in 
   a quantum level due to nonzero fermion condensate 
   $\langle \bar \Psi \Psi \rangle \neq 0$.
The most straightforward demonstration will be the mean field 
   approximation with the concept of self-consistency. 
If one has $N$-component fermion, we can justify 
   the mean field approximation by the leading approximation 
   of $1/N$ expansion.
The self-consistency condition is a crucial key to the 
   description of broken phase. 
This condition directly leads to the gap equation 
 which determines the value of condensate and, equivalently, 
 the physical fermion mass.
As a result of symmetry breaking, there emerges a Nambu-Goldstone boson. 
Since we do not have any fundamental scalar boson in the NJL model, 
 the NG boson (pion) should be supplied dynamically as a bound state 
 of a fermion and an anti-fermion.
Mass of the pionic state indeed vanishes in the chiral limit.

It will be helpful to comment on the physics meaning of treating 
 the model (\ref{aa}). 
First of all, we should clearly distinguish our model from the linear 
 $\sigma$ model of Gell-Mann and L\'evy \cite{Gell-Mann}. 
Structurally our model resembles the $\sigma$ model in the sense that 
  it consists of bosons and fermions interacting with each other 
  via Yukawa couplings and has a continuous chiral symmetry.
However, an important difference in our model is the absence 
  of potential term for scalars.
In the linear $\sigma$ model,   it is the wine-bottle potential that 
  induces the chiral symmetry breaking. 
Therefore the symmetry breaking occurs  in the {\it tree} level
  and the dynamical formation of NG boson can not be seen.
What we obtain is only the sigma condensate $\langle \sigma \rangle\neq 0$.
This naturally leads us to identify the fermions with 
nucleons\footnote{However the fermions are sometimes treated as quarks. 
  For example in the LF formalism, Carlitz et al. 
  \cite{Carlitz} investigated the linear $\sigma$ model 
  (i.e., {\it with} the wine-bottle potential) regarding the 
  fermions as quarks.  But such treatment does not tell 
  anything about {\it dynamical} chiral symmetry breaking 
  and therefore should be clearly distinguished from our standpoint.}. 
On the other hand, the fermions in our model should be regarded 
  as quarks rather than nucleons.
Indeed, as far as the leading order of $1/N$ expansion is concerned,
  the model shows the same behavior as the NJL model. 
For example, straightforward calculations such as effective potential 
  \cite{Kugo} or mean-field approximation show
  that the chiral symmetry breaking $\langle \sigma \rangle = 
  -\frac{\lambda}{N}\langle \bar \Psi \Psi \rangle  \neq 0$ 
  occurs in the one-loop quantum level (see Appendix A for more details).
Therefore we do not consider our model as a special case of the
  linear $\sigma$ model and in order to remind this, 
  we call it the ``chiral Yukawa model''.

It is also very important to view the NJL model as a
  low-energy  effective theory of the chiral Yukawa model. 
The relation between two models  is  very similar to that between 
  the Weinberg-Salam model and the Fermi theory of weak interaction:
The chiral Yukawa model is renormalizable and the fermions interact 
  with each other by exchanging scalar or pseudo-scalar bosons.
If we take the infinitely heavy mass limit for bosons 
  $\mu/\sqrt{\lambda}\rightarrow \infty$, then the theory 
  reduces to the NJL model with non-renormalizable four-Fermi interactions.
In this sense, the NJL model can be considered as a low energy 
  effective theory of the chiral Yukawa model.
Such low energy approximation will be valid when the momentum is much 
  smaller than the mass of the exchanged particles $p^2\ll \mu^2/\lambda$.
In this paper, we mainly treat the chiral Yukawa model for technical reasons, 
  but what we eventually want to know are results of the NJL model. 
Therefore even if we encounter divergences during calculation 
  in the chiral Yukawa model, we only regularize them by some cutoff scheme
  and do not renormalize them.

Finally, as far as we are discussing such 'low energy region', 
   we do not have to worry about the problem of double counting of 
   physics degrees of freedom. 
In the NJL model, the scalar and pseudo-scalar bosons 
  are described as quark-antiquark bound states. 
On the other hand, we treat the scalars as physically independent 
   degrees of freedom  in the chiral Yukawa model. 
So if we regard the fermions and scalars as quarks and mesons, 
   there is the problem of double counting with which we are always 
   confronted in treating chiral-quark type models \cite{ChiralQuark}.
However, in the low energy region, or equivalently for sufficiently large
  boson mass $\mu\rightarrow \infty$, the scalars become ``frozen'' and 
  do not behave as propagating degrees of freedom. 
%

\subsection{Setup in the LF quantization}
We analyze the chiral Yukawa model (\ref{aa}) in the DLCQ method and take 
  special care of the longitudinal zero modes of scalars.
In this method, we compactify the longitudinal space into a circle 
  $x^-\in [-L, L]$ with appropriate boundary conditions on fields.
For scalars, we impose periodic boundary conditions at each 
  LF time 
\begin{eqnarray}
  \sigma(x^- =-L,~ x_\perp) &=& \sigma(x^-= L,~ x_\perp),\\
  \pi(x^-=-L,~ x_\perp) &=& \pi(x^- =L,~ x_\perp),
  \label{bd}
\end{eqnarray}
so that we can explicitly treat the longitudinal zero modes defined by
\begin{eqnarray}
&&\sigma_0(x_\perp)=\frac{1}{2L}\int_{-L}^L dx^-\ \sigma(x),\\
&&\pi_0 (x_\perp)=\frac{1}{2L}\int_{-L}^L dx^-\ \pi(x).
\end{eqnarray}
Then the scalar fields are decomposed into the zero modes and the remaining 
oscillation modes:
\begin{eqnarray}
&&\sigma(x)=\sigma_0(x_\perp)+\varphi_\sigma(x),\\
&&\pi(x)=\pi_0(x_\perp)+\varphi_\pi(x).
\end{eqnarray}

On the other hand, we impose an antiperiodic boundary condition
 for the fermion field,
\begin{equation}
  \Psi^a (x^-=-L,~ x_\perp)= -\Psi^a (x^-=L,~ x_\perp).
  \label{be}
\end{equation}
Here we must be careful about the boundary condition on the 
  ``bad component'' of the fermion.
As we discuss in the next section, if we decompose the fermion as 
\begin{equation}
  \Psi^a = \psi^a_+ + \psi^a_-,\quad  
           \psi^a_{\pm} \equiv \Lambda_{\pm} \Psi^a ,
   \label{abc}
\end{equation}
we find that $\psi_-$ (``bad component'') is a dependent 
  field (see Appendix  for the definition of $\Lambda_{\pm}$).
So the boundary condition on $\psi_-$ should be imposed consistently 
  with the dynamics. 
For example, if we imposed the periodic boundary condition on 
  $\psi_-$ and antiperiodic on $\psi_+$, 
  the mass term $\bar \Psi\Psi$ and the fermion's kinetic term 
  became antiperiodic.
This is not desirable as a term in the lagrangian and even not 
  consistent with the scalar sector.
Then how about the periodic boundary conditions for both of 
  $\psi_+$ and $\psi_-$?
In this case  we have a dynamical zero mode of $\psi_+$, which is,
  however, not important to our problem
  because the chiral condensate will be related to the zero modes 
  of $\sigma$ or $\bar\Psi\Psi$.   
Periodic fermion will give unnecessary intricacy to the problem. 
Therefore the antiperiodic boundary condition (\ref{be}) 
  is appropriate.

\setcounter{equation}{0}

\section{Classical aspect}

Classical analysis is necessary for specifying independent degrees 
  of freedom.  
In this section, we determine the constraint structure of the model 
  and define the LF chiral transformation.
Chiral current and charge are explicitly given.

\subsection{Constraints}
The system has three important constraints characteristic of the 
  LF formalism: a constraint for $\psi_-$ (fermionic constraint) 
  and two constraints for zero modes of bosons (zero-mode constraints).
The Euler-Lagrange equation for $\psi_-$ itself is the 
  fermionic constraint:
\begin{equation}
   i \partial_- \psi^a_- = {1 \over 2}( i\gamma^\perp\partial_\perp
             + m +\sigma - i \pi \gamma_5 )\gamma^+ \psi^a_+ ~.
  \label{acaa}
\end{equation}
Also the zero-mode constraints for $ \sigma_0(x_\perp)$ and 
  $\pi_0(x_\perp)$ are easily obtained  from  $x^-$-integration of 
  the Euler-Lagrange equations for $\sigma$ and $\pi$, respectively: 
$$
\left( \frac{\mu^2}{\lambda} - \partial_\bot^2 \right)
         \left(\matrix{\sigma_0 \cr \pi_0}\right)
      + \frac{\mu^2}{N} \left[
         \bar \Psi^a(x) \left(\matrix{1\cr i\gamma_5}\right)
         \Psi^a(x)\right]_0=0,
$$
where $[\quad]_0$ denotes integration over $x^-$ (see Appendix B).
More explicitly,
\begin{eqnarray}
&&0=\left( \frac{\mu^2}{\lambda} - \partial_\bot^2 \right)
       \left(\matrix{\sigma_0 \cr \pi_0}\right)\nonumber\\
&&     - \frac{\mu^2}{N} \frac{1}{\sqrt{2}} \left[ 
           \psi_+^{a \dag} 
           \left(\matrix{-1\cr i\gamma_5}\right)
           \gamma^- \psi^a_- 
            + \psi_-^{a \dag} 
           \left(\matrix{-1\cr i\gamma_5}\right)
           \gamma^+ \psi^a_+ 
           \right]_0.
  \label{acc}
\end{eqnarray}
These equations mean that $\sigma_0$ and $\pi_0$ should be represented 
   by other independent variables.
If we take the $\mu\rightarrow \infty$ limit, the zero-mode constraints 
   are reduced to zero-mode projected equations of the familiar relations 
   $\sigma=-\frac{\lambda}{N}\bar\Psi\Psi$ and 
   $\pi=-\frac{\lambda}{N}\bar\Psi i \gamma_5\Psi$.     
Eventually the independent degrees of freedom are 
     nonzero modes of the scalars $\varphi_\sigma,\ \varphi_\pi $, and 
     the ``good component'' of the spinor $\psi_+$.
The above constraints are, of course, derived from Dirac's procedure 
  (see Appendix C).
It is easily found that they belong to the second class.

\subsection{Chiral transformation on the LF}

Definition of chiral transformation on the LF is different 
  from the usual one Eqs.~(\ref{ace}) and (\ref{acf}). 
This is because the identification of 
  independent degrees of freedom is not the same as usual.
As we saw,  {\it $\sigma_0, \pi_0$ and $\psi^a_-$ are dependent variables
  and  should change as a result of transformation of the 
  independent variables $\varphi_\sigma, \varphi_\pi$ 
  and $\psi^a_+$.}
Therefore in the LF formulation, we impose the chiral transformation 
only on the dynamical variables:
\begin{equation}
  \psi_+^a   \rightarrow  {\rm e}^{i \theta \gamma_5}  \psi_+^a ~, 
  \label{acg}
\end{equation}
\begin{equation}
   \left( 
      \matrix{ \varphi_\sigma  \cr \varphi_\pi } 
   \right)
      \rightarrow   R(2\theta)
   \left( 
      \matrix{ \varphi_\sigma  \cr \varphi_\pi } 
   \right),
  \label{ach}
\end{equation}
where $R(2\theta)$ represents a rotation matrix defined in Eq.~(\ref{acf}).
These are the definition of the ``LF chiral transformation''.
If we find that $\psi_-$ and $(\sigma_0, \pi_0)$ also transform as 
  $\psi_-\rightarrow {\rm e}^{i\theta\gamma_5}\psi_-$ and 
  $\left(\matrix{\sigma_0\cr\pi_0}\right) \rightarrow R(2\theta)
   \left(\matrix{\sigma_0\cr\pi_0}\right) $ 
  as a result of (\ref{acg}) and (\ref{ach}), 
  we can say that the ``LF chiral transformation'' is substantially 
  equivalent to the usual one (\ref{ace}) and (\ref{acf}).
However, what is surprising about the ``LF chiral transformation''
   is that the transformation (\ref{acg}) is an 
   exact symmetry even for {\it massive} fermion 
   as far as interaction is absent \cite{Mustaki}.
So it will be interesting to check  whether the 
   ``LF chiral transformation'' in our model is exact or not 
   when a mass term is present. 

In order to see the transformation property of the dependent 
  variables, let us solve the constraints {\it classically}.
This means that we completely ignore the ordering of the variables 
  which becomes a burdensome but important issue in a quantum treatment.
The fermionic constraint (\ref{acaa}) which was originally 
  a complicated relation in the purely fermionic NJL model\footnote{It is
  difficult but possible to solve the fermionic constraint in classical 
  treatment where we just treat the spinors as Grassmannian numbers. 
  The exact solution obtained is highly nonlocal and complicated \cite{II}.}, 
  is now easily solved owing to introduction of scalars.
The zero-mode constraints are also solved formally.
Explicit form of the solutions is given in Appendix C.
Now we find the transformation of the zero modes and subsequently 
  that of $\psi_-$. 

\subsubsection{Massless fermion}
Let us first consider the massless fermion case.
When $m=0$, it is easy to see that the transformation (\ref{acg}) 
  and (\ref{ach}) induces the following:
\begin{equation}
     \left(\matrix{\sigma_0 \cr \pi_0} \right)  
     \rightarrow
    R( 2 \theta)
     \left(\matrix{\sigma_0 \cr \pi_0} \right),
  \label{btd}
\end{equation}
\begin{equation}
  \psi^a_- \rightarrow 
  {\rm e}^{i \theta \gamma_5} \psi^a_- .
  \label{bte}
\end{equation}
This is identical with the usual chiral transformation.
Therefore it is shown that when $m=0$ the fields do 
    transform as (\ref{ace}) and  (\ref{acf}) even on the LF 
    at the classical level.

Now that we know all the transformation laws, it is straightforward 
   to construct the Noether current and charge.
The lagrangian with $m = 0$ is
   invariant under the LF chiral transformations.
Form of the LF chiral current is equivalent to the usual one,
\begin{equation}
  j_\mu^5 = - \bar{\Psi}{}^a \gamma_\mu \gamma_5 \Psi^a
            + {2 N \over \mu^2} \Big( \pi \partial_\mu \sigma 
                - \sigma \partial_\mu \pi\Big).
  \label{btf}
\end{equation}
However, $\sigma_0,\ \pi_0$ and $\psi_-$ in (\ref{btf}) should be 
  understood as solutions of the constraints.
On the other hand, the LF chiral charge 
  $Q_5^{\rm LF}= \int_{-L}^L dx^- \int d^2x_\perp \ j^+_5$ 
  does not include the constrained variables:
\begin{equation}
  Q_5^{\rm LF}\!\!  
    =\!\! \int \!\! d^3 x\! \left[-\sqrt{2}\psi_+^{a \dag} \gamma_5 \psi^a_+
       + {2 N \over \mu^2}  \Big(\varphi_\pi \partial_- \varphi_\sigma 
          - \varphi_\sigma \partial_- \varphi_\pi \Big) \right],
  \label{btg}
\end{equation}
which is consistent with the fact that $Q_5^{\rm LF}$ is a 
   generator of the chiral rotation for independent variables.
Transformation of other dependent fields should be obtained 
   through the change of dynamical variables.

\subsubsection{Massive fermion}
The massive fermion case is much more complicated.
As mentioned before, an astonishing fact of the ``LF chiral transformation''
   is that  it is an exact symmetry even for a massive free fermion 
   \cite{Mustaki}.
When the mass term is present, the ``bad component'' of free fermion 
   does not transform as Eq.~(\ref{bte}). 
Subsequently, the associated Noether current $\hat J_\mu^5$ has an extra
 term proportional to the bare mass 
\begin{equation}
\widehat J_\mu^5 = J_\mu^5 - im (\bar\psi_++\bar\psi_-) 
     \gamma_\mu \gamma_5 \frac{1}{\partial_-}\gamma^+\psi_+.
    \label{current}
\end{equation}
Nevertheless, the divergence of the current turns out to be zero 
 $\partial^\mu \widehat J_\mu^5=0$ due to the cancellation between 
  the first term
 ($\partial^\mu J_\mu^5=-2im\bar\Psi\gamma_5\Psi$) and the second term. 
This should be compared with the usual current in the equal-time 
  quantization: ${\cal J}_\mu^5=-\bar\Psi \gamma_\mu \gamma_5 \Psi,\ 
  \partial^\mu {\cal J}_\mu^5=-2im \bar\Psi\gamma_5 \Psi$, 
  which also holds for interacting theories and is intimately 
  connected with PCAC relation.
Note that the LF chiral charge $\widehat Q_5^{\rm LF}= \int d^3x 
  \widehat J^+_5$  is equivalent to that in the massless case due to 
  $\widehat J^+_5 = J^+_5$.
This is natural because the LF chiral transformation is defined irrespective 
  of the mass term.

Now, how about the chiral Yukawa model?
Using the solution for $m\neq 0$ (see Appendix C), 
  the infinitesimal chiral transformation of 
  $\sigma_0$, $\pi_0$ and  $\psi_-$ are given as follows: 
\begin{eqnarray}
   &&  \left(\matrix{\delta \sigma_0 \cr \delta\pi_0} \right)=
     \left(\matrix{2\theta \pi_0\cr -2\theta\sigma_0}\right)
   + \left(\matrix{0\cr 2\theta m \zeta } \right) ,
  \label{massive1}\\
  &&\delta\psi^a_- =
  {i \theta \gamma_5} \psi^a_-
   -2 i \theta \gamma_5 m ( 1+ \zeta ) \frac{1}{2} \frac{1}{i \partial_-}
\gamma^+ \psi_+  \: ,
  \label{massive2}
\end{eqnarray}
where
$$
 \zeta \equiv
   -{1 \over \sqrt{2}}{\mu^2 \over N} {\cal D}^{-1}(x_\perp)
	\left[
	     \psi_+^{a \dag}\frac{1}{i \partial_-}\psi^a_+
	 +\ {\rm c.c.}\
 	\right]_0  ,
$$
and ${\cal D}(x_\perp)$ is the transverse differential operator 
   defined in Appendix C. 
It is evident that the dependent fields do not transform as 
Eqs.~(\ref{btd})
  and (\ref{bte}).
As a result, the Noether current (\ref{btf}) also gets modified by the term 
  proportional to $m$.
The explicit form of the current $\widehat j^\mu_5$ is very complicated 
  but we can see that the $+$ component is equivalent to that of (\ref{btf}).
Therefore the LF chiral charge is given by Eq.~(\ref{btg}) even 
  for massive case.
However, contrary to the massive free fermion, the divergence of the 
  LF current $\partial^\mu \widehat j_\mu^5$ does not vanish due to 
  nontrivial interactions.

One of the lessons suggested by these observations is that when we  
  investigate physics related to massive fermions (e.g. PCAC relation), 
  we had better treat the current $j_\mu^5$ defined by the massless fermions
  rather than $\widehat j_\mu^5$.
This is clear for the free case:
The true LF chiral current $\widehat J_\mu^5$ (\ref{current}) defined 
  for the massive fermion vanishes if we take the divergence 
  while that for massless fermion $J_\mu^5$ gives the usual relation.
This is true of the chiral Yukawa model.
The divergence of the current (\ref{btf}) is given as
\begin{equation}
\partial^\mu j_\mu^5 = -2m \bar\Psi i \gamma_5 \Psi
\label{divergence}
\end{equation}
while $\partial^\mu \widehat j_\mu^5$ is very complicated.
It will be very difficult to discuss the PCAC relations etc. by using 
  $\widehat j_\mu^5$.
It is not quite clear if analysis of $\widehat j_\mu^5$ makes sense.
Instead, even for the massive case, we treat the current (\ref{btf}) 
  to discuss the physics such as PCAC.
This point will be discussed later again in Section V.

\setcounter{equation}{0}
\section{Quantum aspects}

In the classical analysis, we formally solved the constraints  
  in order to find the LF chiral current and charge.
When $m=0$, the resulting hamiltonian is chiral symmetric 
  and we do not have any symmetry breaking term.
Therefore even if we go to quantum theory with such hamiltonian, 
  we will not be able to describe the chiral symmetry breaking.
Certainly it might be possible that we could find a broken phase 
  hamiltonian by adjusting the operator ordering, 
  but such procedure seems unnatural and tricky. 
Instead, we quantize the model before solving the constraints. 
This means that we perform the Dirac quantization for constrained systems.
After that, the constraints are solved quantum mechanically with 
  a care of the operator ordering.
The same route has been traced by many people who tried 
 to describe the spontaneous symmetry breaking of simple scalar systems 
 \cite{SSB}.

Calculation of the Dirac brackets in our system is 
  a very complicated task.
However, the Dirac brackets between dynamical variables turn
  out to be standard ones: Quantization conditions for the dynamical 
  variables $ \varphi_\sigma, \varphi_\pi,$ and $\psi^a_+$ are
\begin{eqnarray}
 &&\Big[\varphi_{\xi} (x),~\varphi_{\eta}(y)\Big]_{x^+=y^+}\label{raa} \\
 &&\quad =- \delta_{\xi \eta} \frac{\mu^2}{N} \frac{i}{4}
     \left( \epsilon( x^- - y^- )-\frac{x^- - y^-}{L}\right) 
        \delta^{(2)} (x_\bot -y_\bot) ,
   \nonumber\\
 &&\left\{ \psi_{+\alpha}^a (x),\psi_{+\beta}^{b \dag}(y) 
    \right\}_{x^+ =y^+}\nonumber\\
 &&\quad = \frac{1}{\sqrt2}(\Lambda_+)_{\alpha \beta} 
           \delta^{ab}\delta (x^- - y^- )
           \delta^{(2)} (x_\bot -y_\bot) ,
  \label{rab}
\end{eqnarray}
where $\xi$ and $\eta$ stand for $\sigma$ or $\pi$, and 
  $\alpha, \beta=1,\cdots,4$ are the spinor indices. 
The sign function $\epsilon(x^-)$ is defined in Appendix B. 
The other commutators between dynamical variables are zero.
Note that these conditions are {\it irrespective of the phase} 
  of the model because they are independent of the interaction.

Mode expansion of the fields at $x^+ =0$ reads 
\begin{eqnarray}
&&\varphi_\eta({\bf x})= \sqrt{\frac{\mu^2}{N}}
\frac{1}{2L}\sum_{n=1,2 \cdots}\frac{1}{2p_n^+}\int \frac{d^2p_\perp}{(2\pi)^2}
\nonumber\\
&&\hskip1cm \times \left\{  a_\eta (p_n^+,{\bf p}_\perp)\ {\rm e}^{-i{\bf px}}
       + a_\eta^{\dag}(p_n^+,{\bf p}_\perp)\ {\rm e}^{i{\bf px}}
\right\}, \\
&&\psi_+^a({\bf x})=
 \frac{1}{2L}\sum_{n=\frac12,\frac32 \cdots}
   \int \frac{d^2p_\perp}{\sqrt{2\pi}2^{3/4}\sqrt{p_n^+}}\nonumber \\
&&\hskip2cm \times \sum_{h=\pm \frac12}\left\{
    w(h)b^a(p_n^+,{\bf p}_\perp,h)\ {\rm e}^{-i{\bf px}}\right. \nonumber\\
&&\hskip3cm  + \left. w(-h)d^{a\dag} (p_n^+,{\bf p}_\perp, h)\ 
     {\rm e}^{ i{\bf px}}  \right\},
\end{eqnarray} 
where ${\bf px}=p_n^+x^--{\bf p}_\perp {\bf x}_\perp$ and 
  $p_n^+=\pi n/L$. 
The spinors $w(\pm h)$ depend only on the helicity $h$ \cite{Mustaki}.
It follows that
\begin{eqnarray}
 && \left[ a_\xi (p_n^+,{\bf p}_\perp),\ 
           a^\dag_\eta(q_m^+,{\bf q}_\perp)\right]\nonumber\\
 &&\hskip3cm 
   =(2\pi)^2 (2L) 2p_n^+
     \delta_{n m}\delta({\bf p}_\perp-{\bf q}_\perp) \delta_{\xi \eta}, 
    \nonumber\\
 && \left\{ b^a(p_n^+,{\bf p}_\perp,h),\ 
            b^{b\dag}(q_m^+,{\bf q}_\perp,h')\right\}\nonumber\\
 &&\hskip3cm =2p_n^+  \frac{2L}{2\pi} \delta_{n m} 
              \delta({\bf p}_\perp-{\bf q}_\perp)\delta_{hh'},\nonumber\\
 && \left\{ d^a(p_n^+,{\bf p}_\perp,h),\ 
            d^{b\dag}(q_m^+,{\bf q}_\perp,h')\right\}\nonumber\\
 &&\hskip3cm =2p_n^+  \frac{2L}{2\pi} \delta_{n m} 
             \delta({\bf p}_\perp-{\bf q}_\perp)\delta_{hh'}.\nonumber
\end{eqnarray}
It is important to note that both of the above mode expansions are 
  independent of the mass.
(The spinors $w(\pm h)$ are independent of mass. This 
  is clearly shown in the Appendix of Ref.~\cite{Heinzl_review}.)
This means that if we calculate $n$-point Green functions at fixed 
  time $x^+=0$, they will become independent of the value of mass, which is 
  not a correct result in general. 
This undesirable situation is known as one of the pathological properties 
  of the LF formalism  which needs great care for obtaining correct 
  results \cite{Naka-Yama}. 
Indeed, as we will see later, to remedy this problem is indispensable 
   to get a meaningful gap equation.
In many cases, loss of mass information is cured by a carefully chosen 
   infrared regularization.
It should also be commented that the mass-information loss  
   is certainly not a desirable feature, but we will find its usefulness 
   in describing the broken phase physics.
Anyway, we must pay great attention to the fact that naive mode expansion 
 of the  (scalar and fermion) fields is independent of the value of mass.

The Fock vacuum $|0\rangle$ is  defined by
\begin{eqnarray}
&&a_\sigma (p_n^+,{\bf p}_\perp)|0\rangle
= a_\pi (p_n^+,{\bf p}_\perp)|0\rangle\nonumber \\
&&\quad =b^a(p_n^+,{\bf p}_\perp,h)|0\rangle 
=d^a(p_n^+,{\bf p}_\perp,h)|0\rangle=  0, 
\label{vac}
\end{eqnarray}
for $n>0$. 
It is worth while emphasizing that the vacuum of this system 
  is really the Fock vacuum since we have no dynamical zero modes. 
Because of the $p^+$ conservation, the normal-ordered chiral charge
$Q_5^{\rm LF}$ {\it always} annihilates the vacuum:
\begin{equation}
 Q_5^{\rm LF} |0\rangle= 0.
\end{equation}
It has been known that any light-like  charge $Q^{\rm LF}$ automatically 
  leaves the vacuum invariant  $Q^{\rm LF} |0\rangle= 0$ 
  whether or not it generates a symmetry \cite{Leutwyler}.

In a quantum theory, operator ordering becomes an issue.
Let us comment on the problem of 
  operator ordering and clarify our stance toward it.
Since the (zero-mode and fermionic) constraint equations are 
  generally nonlinear relations among operators, 
  their solutions depend on operator ordering.
We must select an appropriate operator ordering.
Then, what can be the criterion for this problem?
In many papers discussing the spontaneous symmetry breaking in DLCQ, 
  the Weyl ordering is adopted on general grounds. 
However, it is not clear whether the Weyl ordering in constraint 
  equations makes sense because they include both independent and 
  dependent variables.
The most reliable criterion for determining the operator ordering 
  will be as follows.
Before solving the constraint equations, we can calculate the Dirac 
  brackets between independent variables and  dependent ones 
  ({\it e.g.} $[\sigma_0, \psi_+]=\cdots$), 
  which are terribly complicated in our model and we do not display 
  them in this paper. 
Here we already have to specify the operator ordering.
On the other hand, we can solve the constraint relations 
  with the above ordering and obtain their solutions such as 
  $\sigma_0=\sigma_0(\varphi_\sigma, \varphi_\pi, \psi_+)$.
Now we can calculate again the commutators between the solutions 
  (i.e. dependent variables) and independent variables 
  ({\it e.g.} $[\sigma_0(\varphi_\sigma,\varphi_\pi, \psi_+), \psi_+]$ ) 
  using simple commutators Eqs.~(\ref{raa}) and (\ref{rab}).
The results should be identical with those of the Dirac bracket.
In other words, we must find out such operator ordering that will 
  give a consistent result in the above sense.
This should be the criterion for an appropriate operator ordering.
However, as you expect,  to find such ordering in our model is 
  an extremely difficult task. 
So practically,  we  just work with several particular
  orderings and compare the results.  
In our actual calculations, we treat two specific orderings and
  check whether the results depend on the ordering or not.
To find a consistent operator ordering should be examined in 
  much simpler models.

In the following, we will solve the zero-mode constraints 
  in two different ways: perturbative and non-perturbative methods.
To solve the constraint is significant to describe the symmetry 
  breaking on the LF.
To see this, let us decompose the longitudinal zero modes into 
  c-number parts and normal-ordered operator parts,
\begin{eqnarray}
 && \sigma_0 = \sigma_0^{({\rm c})} + \sigma_0^{({\rm op})} ,
  \label{cae}\\
 && \pi_0 = \pi_0^{({\rm c})} + \pi_0^{({\rm op})} .
  \label{caeb}
\end{eqnarray}
If the c-number part of the solution is nonvanishing, 
   it directly means nonzero condensate: 
   $\langle 0 | \sigma | 0\rangle =\sigma_0^{({\rm c})}\neq 0$ and 
   $\langle 0 | \pi | 0\rangle =\pi_0^{({\rm c})}\neq 0 $.
Therefore to find such nontrivial solution is necessary to describe 
  the symmetry breaking.
We explicitly demonstrate that  perturbative solutions cannot 
  lead to chiral symmetry breaking while nonperturbative solutions 
  give nonzero vacuum expectation value for $\sigma$.  
In both cases, the fermionic constraint is formally solved (as in 
  Eq.~(\ref{acaab})) and inserted into the zero-mode constraints.

\subsection{Perturbative solutions to the zero-mode constraints}

Let us  solve the zero-mode constraints using perturbation 
  in terms of the coupling constant $\lambda$.
Since $\lambda$ is a dimensionful parameter, 
  we introduce some scale $\lambda_{\rm cr}$ which is much larger
  than $\lambda$ ($\lambda_{\rm cr}\gg \lambda$). 
We regard $\lambda_{\rm cr}$ as a critical coupling of the 
  symmetry breaking which will be determined later. 
Now we expand the constrained variables as follows
\begin{eqnarray}
   &&\sigma_0 = \sum_{n=0}^{\infty}
     \left( \frac{\lambda}{\lambda_{\rm cr}} \right)^n\sigma_0^{(n)}, 
    \label{exp1}\\
   &&\pi_0 = \sum_{n=0}^{\infty} 
     \left( \frac{\lambda}{\lambda_{\rm cr}} \right)^n\pi_0^{(n)},
    \label{exp2}\\
   &&\psi^a_- = \sum_{n=0}^{\infty} 
    \left( \frac{\lambda}{\lambda_{\rm cr}} \right)^n\psi_{-}^{a (n)},
  \label{exp3}
\end{eqnarray}
and the dynamical variables are treated as 
   $O((\frac{\lambda}{\lambda_{\rm cr}})^0)$.
Inserting the above expansions into the constraints 
 Eqs.~(\ref{acaa}) and (\ref{acc}) with  the natural ordering
   and comparing the same order of perturbation, 
   we obtain the solution order by order.
For example, the lowest order solutions are 
\begin{eqnarray}
   &&\sigma_0^{(0)}= \pi_0^{(0)} =0,\\
  \label{bad}
   &&\psi_-^{a (0)}={1 \over  2}{1 \over i \partial_-} 
    ( i \gamma^\perp \partial_\perp + m + \varphi_\sigma - i \varphi_\pi \gamma_5 ) \gamma^+ \psi^a_+ .
  \label{bab}
\end{eqnarray}
Higher order solutions are given in Appendix D.

The chiral transformation of the perturbative solution in the massless case
 can be inductively checked.
First, it is easy to see that the 0-th and 1st order solutions rotate 
   symmetrically under the LF chiral transformation (\ref{acg}) and 
   (\ref{ach}).
If we suppose the $n$-th order solutions rotate chirally, 
  the $n+1$-th order solutions also behave the same.
Eventually the perturbative solution transforms symmetrically 
   under the LF chiral rotation (\ref{acg}) and (\ref{ach}),
\begin{eqnarray}
&&  \left[ Q_5^{\rm LF} , \Psi^a \right]= \gamma_5 \Psi^a ~,
  \label{bag}\\
&& \left[ Q_5^{\rm LF} , \sigma \right]= -2 i \pi ~, \qquad
   \left[ Q_5^{\rm LF} , \pi \right]= 2 i \sigma ~.
  \label{bah}
\end{eqnarray}
Of course the hamiltonian  [cf) Eq.~(\ref{ham})] is also invariant 
 $[Q_5^{\rm LF},\ H]=0$ 
which is the same as the classical analysis in the previous section.

What is most important is that {\it the
  vacuum expectation values of the perturbative solutions
  vanish in all order of perturbation} $\langle 0 | \sigma_0 | 0 \rangle  
  =  \langle 0 | \pi_0 | 0 \rangle = 0$. 
This is easily verified by using 
   $\langle 0| \varphi_{\sigma, \pi} | 0\rangle =0$ and
   $\langle 0| \psi_+^\dag \partial_\perp \psi_+ |0\rangle=0$.
Therefore we are in a chiral symmetric phase:
\begin{equation}
    \langle 0 \vert \sigma \vert 0 \rangle  
 =  \langle 0 \vert \pi \vert 0 \rangle=0 .
  \label{tys}
\end{equation}

\subsection{Nonperturbative solutions to the zero-mode constraints}

We next solve the zero-mode constraints using the mean-field 
  approximation.
In the following, we work with a particular operator ordering though 
  the result is the same as others as far as we discuss the 
  leading order of $1/N$ expansion.
The following ordering greatly reduces our calculation.
Substituting the solution of the fermionic constraint into the 
  zero-mode constraints and rearrange the ordering, we obtain 
\begin{eqnarray}
 0 &= & \Big({\mu^2 \over \lambda}-\partial_\perp^2 \Big) 
      \left(\matrix{\sigma_0\cr \pi_0} \right)\nonumber\\
  &+& {1 \over 2^{3/2}} {\mu^2 \over N}{1 \over 2 L} 
      \int_{-L}^{L}\!\! dx^- \int_{-L}^{L}\!\! dy^-
      \frac{\epsilon (x^- -y^-)}{2i}          \nonumber  \\ 
  &\times&  \left\{
        i \psi_+^{a \dag}(x)\left(\matrix{-1\cr i\gamma_5}\right) 
        \gamma^\perp \partial_\perp \psi^a_+(y)\right.\nonumber\\
  &&\quad     -i \partial_\perp \psi_+^{a \dag}(y) 
        \left(\matrix{-1\cr i\gamma_5}\right)\gamma^\perp\psi^a_+(x)
                                            \nonumber  \\
  &+ &  \left(\matrix{m+\sigma(y)\cr \pi(y)} \right) 
       \left( \psi_+^{a \dag}(x) \psi^a_+(y)
                       - \psi_+^{a \dag}(y) \psi^a_+(x)\right)\nonumber  \\
  &-&  \left. \left(\matrix{-\pi(y)\cr m+\sigma(y)}\right) 
       \left( \psi_+^{a \dag}(x) i\gamma_5\psi^a_+(y) 
       + \psi_+^{a \dag}(y)i\gamma_5\psi^a_+(x)
       \right)
       \right\} \nonumber\\
  & &       +\ {\rm h.c.},
  \label{caa}
\end{eqnarray}
where $\psi^a_+(y)=\psi^a_+(x^+,y^-,x_\perp)$ and similarly for $\sigma(y)$ 
  and $\pi(y)$
The operator ordering here is different from that in the perturbative 
   treatment.
However, one can show that the previous perturbative result does not 
  change with the above ordering.
That is, the {\it perturbative} solution with the above ordering does not 
  lead to chiral symmetry breaking.

Let us first determine the c-number part of the zero modes defined by 
  Eqs.~(\ref{cae}) and (\ref{caeb}). 
We saw in the classical analysis that $(\sigma_0, \pi_0)$ rotates chirally
 in the massless case. Therefore we choose 
\begin{eqnarray}
  &&\sigma_0^{({\rm c})} \neq 0 ,
  \label{caf}\\
  &&\pi_0^{({\rm c})} = 0 .
  \label{cag}
\end{eqnarray}
Taking a vacuum expectation value of the zero-mode constraint
  for $\sigma$ greatly 
  simplifies the calculation, which is an advantage of 
  our specific choice of ordering: 
\begin{eqnarray}
  \frac{\mu^2}{\lambda} \sigma_0^{(\rm c)}
 & &\!\!\!\! = -\frac{1}{\sqrt{2}} \frac{\mu^2}{N} 
       (m + \sigma_0^{(\rm c)} )\nonumber\\
&\times&\left\langle 0\left|
         \left[\psi_+^{a \dag}\frac{1}{i \partial_-}\psi^a_+
         - \left( \frac{1}{i \partial_-} \psi_+^{a \dag} \right) \psi^a_+
    \right]_0 \right| 0\right\rangle .
  \label{cah}
\end{eqnarray}
Introducing $M$ defined by
\begin{equation}
  M \equiv m +\sigma_0^{(\rm c)}~,
\end{equation}
and evaluating the vacuum expectation value in Eq.~(\ref{cah}) 
  by using the mode expansion, we have 
\begin{eqnarray}
  M&-&m\nonumber\\
 &=& -\frac{1}{\sqrt{2}} \frac{\lambda}{N} M
    \left\langle 0\left| \left[\psi_+^{a \dag}\frac{1}{i \partial_-}\psi^a_+
         - \left( \frac{1}{i \partial_-} \psi_+^{a \dag} \right) \psi^a_+
    \right]_0   \right| 0\right\rangle \nonumber  \\
  &=& \lambda M {2 \over (2 \pi)^3 }\int d^2 p_\bot \sum_{n=1/2,\cdots} 
       { \Delta p^+ \over p_n^+}  ,
  \label{cai}
\end{eqnarray}
where $\Delta p^+=\pi/L$.
The operator form of the right-hand-side suggests us to identify  as
\begin{eqnarray}
&&\frac{M}{\sqrt2}
    \left\langle 0\left| \left[\psi_+^{a \dag}\frac{1}{i \partial_-}\psi^a_+
         - \left( \frac{1}{i \partial_-} \psi_+^{a \dag} \right) \psi^a_+
    \right]_0   \right| 0\right\rangle\nonumber\\
&&=\left\langle 0 \left| \left[\bar \Psi_{M}^a \Psi_{M}^a \right]_0 
    \right| 0\right\rangle ,
\label{identify}
\end{eqnarray}
where $\Psi_{M}^a$ is a fermion with mass $M$,
\begin{eqnarray}
    \Psi^a_{M} &=& \psi^a_+ + \psi^a_{-M}~,  \nonumber  \\
    \psi^a_{-M} &\equiv& {1 \over 2 } {1 \over i \partial_- }
      ( i \gamma^\perp \partial_\perp + M ) \gamma^+ \psi^a_+ ~.
  \label{cec}
\end{eqnarray}
($\psi^a_{-M}$ is the ``bad'' component of $\Psi^a_{M}$.)
Therefore it is natural to consider $M$ to be the physical fermion mass.
In other words, the identification in Eq.~(\ref{identify}) corresponds to 
  the self-consistency condition.

Eq.~(\ref{cai}) should be the gap equation by which we can 
   determine $\sigma_0^{(c)}$ and equivalently, 
   the physical fermion mass $M$. 
However, it is not evident to regard it as the gap equation
   because Eq.~(\ref{cai}) in the chiral limit $m\rightarrow 0$ cannot 
   give nonzero $M$.
The same situation was observed in our previous work on the Gross-Neveu model 
   \cite{ItakuraMaedan}.
As was discussed in Ref.~\cite{ItakuraMaedan},
   if we want a meaningful gap equation, we must supply mass information 
   so that Eq.~(\ref{cai}) possesses a nontrivial solution $M\neq 0$ 
   in the chiral limit 
   when we regularize the divergent summation over $n$. 
The need of the mass dependence in Eq.~(\ref{cai}) is readily understood 
   from the identification in Eq.~(\ref{identify}).   
Indeed, one can easily check that 
   $\langle 0 | \bar\Psi_M\Psi_M|0\rangle/M $ should explicitly 
   depend on $M$ in the equal-time formulation.
This is a typical example of the ``mass-information loss'' on the LF 
   \cite{Naka-Yama}
   which must be repaired properly for obtaining correct results.

It may be possible to regularize the divergent summation in 
   Eq.~(\ref{cai}) with, say, a heat-kernel damping factor 
   \cite{ItakuraMaedan}, 
   but such calculation is complicated and 
   not tractable.
Instead, we introduce some cutoff that renders the divergent summation 
   into finite one.
Such cutoff should be introduced so that the result correctly depends 
   on the mass $M$.
Here for simplicity, we adopt a cutoff which eventually reduces 
   to the parity invariant (PI) 
   cutoff $p^\pm < \Lambda$ \cite{Itakura}.
 From the dispersion relation and the PI cutoff, we find that 
   the momentum region is restricted to
   $ (M^2 +{\bf p}_\bot^2)/2\Lambda < p_{n}^{+} =\pi n/L < \Lambda $.
Therefore we set
\begin{equation}
  n_{\rm IR} < n < n_{\rm UV} ,
\end{equation}
where $n_{\rm IR}$ and $n_{\rm UV}$ are nearest half-integers to
   $ \frac{L}{\pi} \frac{M^2 +{\bf p}_\bot^2}{2\Lambda}$ 
   and $ \frac{L}{\pi} \Lambda$ respectively.
If we use the approximation for a large half-integer $\widetilde n$,
$ \sum_{n=1/2,\cdots}^{\widetilde n}n^{-1} \simeq {\rm ln}{\widetilde n}
  + {\rm ln}4e^\gamma $,
  the summation is approximated as
\begin{equation}
  \sum_{n=1/2,\cdots}{ \Delta p^+ \over p_n^+} 
  = \sum_{n=n_{\rm IR}+1}^{n_{\rm UV}-1} \frac{1}{n} 
  \simeq {\rm ln} \frac{n_{\rm UV}}{n_{\rm IR}}
  \simeq {\rm ln} \frac{2\Lambda^2}{M^2 +{\bf p}_\bot^2} ,
\end{equation}
for fixed $\Lambda$, $M$ and ${\bf p}_\bot^2$ and sufficiently large $L$.
Of course there is ``finite volume effect'' for finite $L$, 
   but we finally take the infinite volume 
   limit and the finite volume effect is expected to be small
   as far as $L$ is large enough\footnote{
Finite volume physics itself is  intriguing. 
For example, similarly to  the equal-time calculation \cite{Finite}, 
  if we make the volume smaller and smaller,  
  we will meet a critical length $L_{\rm cr}$ beyond which the 
  chiral symmetry never breaks down.
Moreover, if we could determine the $L$ dependence of 
  the physical mass $M(L)$, it would serve as a prediction for 
  the limiting behavior of eigenvalues in the numerical DLCQ calculation.
Nevertheless, such finite volume physics is outside the scope of 
  this paper and we do not discuss it anymore.  
We always assume $L$ sufficiently large and ignore the finite volume effects.
}.
Eventually  Eq.~(\ref{cai}) becomes {\it dependent} on the mass $M$ and 
   can be considered to be a gap equation 
\begin{equation}
  M-m = \lambda M \frac{\Lambda^2}{4 \pi^2}
         \left\{ 2- \frac{M^2}{\Lambda^2} \left( 
            1+{\rm ln}\frac{2 \Lambda^2}{M^2}
                \right) \right\}  .
  \label{caic}
\end{equation}
Form of this gap equation is different from those with familiar cutoff 
  schemes such as the three or four momentum cutoff \cite{NJL_review}, 
  but our gap equation behaves exactly the same as usual.
Indeed, even in the chiral limit $m \rightarrow 0$, this equation 
  is a nonlinear equation for $M$ and when
  the coupling constant $\lambda$ is larger than the 
  critical value $\lambda_{\rm cr}=2 \pi^2/\Lambda^2$, 
  there is a nontrivial solution $ M= M_0\neq 0$  (see FIG.~1). 
This also means that the zero mode of $\sigma$  has been determined as 
  $\sigma_0^{({\rm c})} =M-m$.
Furthermore, one should note that the gap equation (\ref{caic}) 
  is independent of the value of $\mu$.
So we can regard the finite $\mu$  result 
  $\langle \sigma \rangle =M-m$ as the result for infinite $\mu$; 
  $-\frac{\lambda}{N}\langle \bar\Psi\Psi \rangle =M-m$. 
(Remember that the $\mu \rightarrow \infty$ limit of (\ref{acc}) is 
  $\sigma_0=-\frac{\lambda}{N}[\bar \Psi\Psi]_0$.)
Therefore the chiral symmetry breaking occurs for arbitrary value 
  of $\mu$ in the mean-field approximation.
This is consistent with the result of the conventional 
  equal-time quantization (see Appendix A).

On the other hand, there is a trivial solution $M=0$ (when $m=0$) 
   even for $\lambda > \lambda_{\rm cr}$ and if we select this solution
   the resulting theory becomes chiral symmetric.
Then there comes a problem which solution should be physically realized.
Unfortunately, comparison of the vacuum energy for both phases 
   does not tell anything about this problem because the vacuum 
   energies turn out to be the same.
If we found the consistent operator ordering as discussed before, 
   we could estimate difference of the vacuum energies and determine 
   the physically realized phase.
Even without such calculation, however, we can say that the 
   symmetric solution is excluded for $\lambda > \lambda_{\rm cr}$.
This is because there emerge tachyonic modes and the system 
   becomes unstable if we select a trivial solution for 
   $\lambda > \lambda_{\rm cr}$.
This will be again discussed in Section V-B.
So we deal with only the nontrivial solution for 
   $\lambda > \lambda_{\rm cr}$ and do not consider the symmetric solution.

\begin{figure}[hbt]
\begin{center}
\vspace*{1cm}
\psfig{file=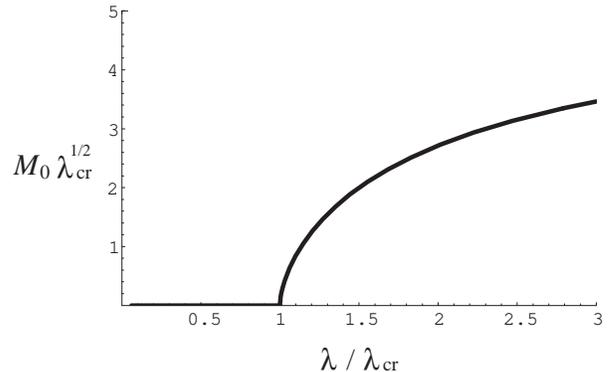, height=5cm, width=8cm}
\caption[]{Fermion's dynamical mass $M_0$ in the chiral limit 
           as the solution of the gap equation (\ref{caic}). 
       There is a nonzero solution for $\lambda > \lambda_{\rm cr}$.}
\end{center}
\end{figure}

Comments on other cutoff schemes are in order. 
We find a nontrivial equation for $M$ by using the PI cutoff.
It was crucial to include the mass information as the regularization.
However, we have to be careful in setting the cutoff.
Any cutoff scheme which holds mass does not necessarily lead to 
   a physically sensible result \cite{Heinzl_cutoff}. 
For example, a two dimensional PI cutoff $M^2/2\Lambda < p^+ < \Lambda$ 
   with a transverse cutoff $|{\bf p}_\perp|<\Lambda$ gives a wrong result.
The resulting gap equation erroneously predicts that there is no symmetric 
  phase. 
It seems important to introduce a cutoff with some symmetry considerations.
Indeed, the three momentum cutoff respecting the three dimensional rotation 
  \cite{LFNJL} and the PI cutoff \cite{Itakura} 
  in our case predict the existence of the critical coupling constant.

Now let us determine the operator parts of the zero modes 
   by the mean-field approximation. 
We approximate the nonlinear terms in Eq.~(\ref{caa})
   by using  $AB\approx A\langle B\rangle + 
   \langle A \rangle B - \langle A \rangle\langle B\rangle$, 
   where the expectation values are taken with respect to the Fock vacuum.
We further neglect contribution from the oscillating modes of scalars
 $\varphi_\sigma=\varphi_\pi =0$.
Then the operator parts  are given by
\begin{equation}
\left(\matrix{\sigma^{(\rm op)}_0\cr\pi^{(\rm op)}_0}\right)
 = -{\mu^2 \over N} \left( m_{\rm ZM}^2 - \partial_\bot^2 \right)^{-1} 
   \left[ : \bar \Psi^a_{M}\left(\matrix{1\cr i\gamma_5}\right)  
          \Psi^a_{M} : \right]_0 ,
  \label{cea}
\end{equation}
where a c-number quantity $m_{\rm ZM}^2$ is defined as
\begin{eqnarray}
&&    m_{\rm ZM}^2 = 
    \frac{\mu^2}{\lambda}\label{ZM_mass}\\
&&\quad +{\mu^2 \over N}{1 \over \sqrt{2}}
    \left\langle 0 \left| \left[ 
    \psi_+^{a \dag}~ \frac{1}{i \partial_-} \psi^a_+
        - \left( \frac{1}{i \partial_-} \psi_+^{a \dag} \right) \psi^a_+
    \right]_0
           \right| 0 \right\rangle  ~.
  \nonumber
\end{eqnarray}
The numerical value of $m_{\rm ZM}^2$ is calculated if we utilize the 
  gap equation:
  $m_{\rm ZM}^2=\frac{\mu^2 m}{\lambda M}$. 
Inserting the c-number and operator parts of $\sigma_0$ and $\pi_0$ into 
  the solution of the fermionic constraint, we have 
\begin{equation}
\psi_-=\psi_{-M}+\frac{1}{i\partial_-}\frac12 
 \left(\sigma_0^{\rm (op)}-i\pi_0^{\rm (op)}\gamma_5\right)\gamma^+\psi^a_+,
\label{FC}
\end{equation}
where $\psi_{-M}$ is given in Eq.~(\ref{cec}).

To understand what we did above, let us consider the relation 
   between our operator ordering and the $1/N$ expansion.
We have obtained an equation for the c-number part of $\sigma_0$
   (the gap equation) just by taking the vacuum expectation value of 
   the zero-mode constraint even without recourse to the $1/N$ expansion.
This simplicity in obtaining the gap equation is mainly due to 
   our specific choice of the operator ordering.
As has been commented before, if we take other orderings, 
  our calculation becomes terrible because of complicated structure 
  of the Dirac brackets between constrained zero modes and physical variables. 
However, as far as the leading term of the $1/N$ expansion is concerned, 
  the commutator $[\sigma_0, \psi_+]$ turns out to be of the order of 
  $O(1/N)$ and we can ignore the effect of ordering\footnote{If one takes 
  $\mu\rightarrow \infty$, one will be convinced that
   $\left[\sigma_0, \psi_+^a\right] 
   =\left[-\frac{\lambda}{N}[\bar\Psi \Psi ]_0, \psi_+^a\right]
    = O(N^{-1})$.}.
Furthermore, the approximation neglecting the scalar oscillating modes 
  is also justified by the $1/N$ expansion. 
 From the quantization condition (\ref{raa}), 
  we find $\varphi_\sigma$ is $O(N^{-1/2})$ whereas $\sigma_0$ is $O(N^0)$. 
These considerations justify that our mean-field calculation with 
  the specific operator ordering is correct
  up to the leading contribution of the $1/N$ expansion.

Before ending this section, it will be better to point out
 the ``merit'' of the mass-information loss.
Certainly it was a demerit in deriving the gap equation,
 but this property gives a very important benefit to our framework.
The fact that the mode expansion is independent of the value of mass 
  in turn means that the Fock vacuum defined by (\ref{vac}) keeps 
  invariant even if we  change the value of mass.
We do not have to perform the Bogoliubov transformation on the vacuum 
  depending on the change of mass. 
Therefore the LF vacuum is invariant even after the fermion acquires 
  dynamical mass $M\neq 0$.

\setcounter{equation}{0}

\section{Physics in nonperturbative region}
In this section, we discuss some physics consequences of our method. 
Firstly, we explicitly demonstrate 
  how the triviality of the null-plane chiral charge $Q_5^{\rm LF}$ 
  and the nonzero chiral condensate reconcile with each other. 
Secondly, masses of the scalar and pseudoscalar bosons are calculated 
  from the lagrangian for both phases.
Finally, we derive the PCAC relation for the chiral current $j_5^\mu$ 
  [Eq.~(\ref{btf})] and discuss 
  the nonconservation of $Q_5^{\rm LF}$.

\subsection{Null-plane chiral charge vs  chiral condensate}
In the equal-time quantization, the broken vacuum does not possess 
  the chiral symmetry $Q_5^{\rm ET}|0 \rangle\neq 0$. 
The Nambu-Goldstone phase is characterized by a nonzero condensate
  of the order parameter $\langle\bar\Psi\Psi\rangle$ and the more 
  strict expression of $Q_5^{\rm ET}|0 \rangle\neq 0$ is 
  a relation $\langle 0|[Q_5^{\rm ET}, \bar\Psi i \gamma_5\Psi]|0\rangle 
  =\langle 0|\int d^3x\ [j_5^0(x), \bar\Psi i \gamma_5\Psi]|0\rangle
  =2i\langle 0|\bar\Psi\Psi |0\rangle\neq 0$.
Therefore there is no inconsistency  between these two relations.
On the other hand, remember that the light-like charge always annihilates 
  the vacuum. 
This implies that if a similar relation 
 $[Q_5^{\rm (LF)}, \bar\Psi i \gamma_5\Psi]\stackrel{?}{=}2i\bar\Psi\Psi$ 
 held on the light front in the broken phase 
 $\langle 0|\bar\Psi\Psi |0\rangle\neq 0$, 
 it would immediately conflict with the triviality of the chiral charge 
 $Q_5^{\rm LF}|0\rangle =0$.
In the following we resolve this seemingly inconsistent situation.

The chiral charge $Q_5^{\rm LF}$ defined by Eq.~(\ref{btg}) 
  annihilates the vacuum and generates 
  the chiral transformation for the independent variables 
  irrespective of the symmetry.
Indeed, we find  
\begin{eqnarray}
 && \left[ Q_5^{\rm LF} , \varphi_\sigma \right]=-2i \varphi_\pi,\quad 
    \left[ Q_5^{\rm LF} , \varphi_\pi \right]= 2i \varphi_\sigma, \\
 && \left[ Q_5^{\rm LF} , \psi_+ \right]=  \gamma_5 \psi_+ .
\end{eqnarray}
These are the fundamental laws of the chiral transformation. 
Any transformation of the dependent fields 
  $\sigma_0,\ \pi_0$, and $\psi_-$ should be derived from them.

In the broken phase $ \lambda > \lambda_{\rm cr}$, the gap equation has
   a nontrivial solution $M\neq 0$ and the fermion behaves as 
   a massive fermion with the dynamical mass $M$.
First of all, let us view the chiral transformation of the massive 
   fermion operator $\Psi_M$ defined by Eq.~(\ref{cec}).
The result is already unfamiliar to us:
\begin{eqnarray}
&&  \left[ Q_5^{\rm LF} , \Psi^a_{M} \right] 
  = \gamma_5 \Psi^a_{M} +\Delta \Psi_M^a,
  \label{t}\\
&&\Delta \Psi_M^a\equiv - 2M \gamma_5 
      \frac{1}{2} \frac{1}{i \partial_-} \gamma^+ \psi^a_+. 
\end{eqnarray}
The second term $\Delta \Psi_M^a$ does not exist in the equal-time 
  quantization.
Only if $M\neq 0$, this is equivalent to the usual transformation.
Using this result, the chiral transformation of $\sigma_0^{(\rm op)}$,
  $\pi^{(\rm op)}_0$ in Eq.~(\ref{cea}) and $\psi_-$ in Eq.~(\ref{FC})
  are given as follows:
\begin{eqnarray}
  &&  \left[ Q_5^{\rm LF} ,\ \sigma_0^{(\rm op)} \right]
        = -2 i \pi_0^{(\rm op)} +\Delta \sigma_0^{\rm (op)},
  \label{bawb}       \\
  &&\left[  Q_5^{\rm LF} ,\ \pi_0^{(\rm op)} \right]
       =   2 i \sigma_0^{(\rm op)} +\Delta \pi_0^{\rm (op)},
  \label{bawa}      \\
  &&\left[ Q_5^{\rm LF} ,\ \psi^a_- \right] =
        \gamma_5 \psi^a_- +\left(1+\frac{i}{2M}\Delta\pi_0^{\rm (op)}\right)
        \Delta\Psi_M^a,
\end{eqnarray}
where
\begin{eqnarray}
   \left( \matrix{ \Delta \sigma_0^{\rm (op)} \cr \Delta \pi_0^{\rm (op)} }
   \right)
  &=& 2i\frac{\mu^2}{N} \frac{1}{m_{\rm ZM}^2-\partial_\bot^2}
    \left\{ \langle 0|\bar \Psi^a_{M} 
            \left( \matrix{ i\gamma_5 \cr -1}\right) \Psi^a_{M}|0 \rangle 
    \right.
    \nonumber\\
&&    \left.   + \left[\frac{M}{\sqrt2} \psi_+^{a \dag} 
                        \left(\matrix{ i\gamma_5 \cr 1}\right)
                       {1 \over i\partial_-}
                  \psi^a_+ + {\rm h.c.}    
            \right]_0 
   \right\} .\nonumber
\end{eqnarray}
One can easily show that $\Delta \sigma_0^{\rm (op)}$ is zero due to 
  $\langle 0 | \bar \Psi_M i \gamma_5 \Psi_M | 0 \rangle =0$ and 
  antisymmetry of the sign function $\epsilon(x^--y^-)$.
Therefore the terms involving $\Delta\Psi_M$ and $\Delta\pi_0^{\rm (op)}$
  are the extra compared with the usual chiral transformation.
They do not vanish even in the chiral limit.
These extra terms are direct consequences of being dependent variables
  and the dynamical generation of the fermion mass $M$.
Unlike the equal-time calculation, the chiral transformation of the 
  full field variables becomes {\it model-dependent} in general because 
  a part of the variables are constrained and the information of 
  interaction inevitably enters the transformation law of constrained 
  variables through the solutions.

Due to the modification of the transformation law, 
   we can avoid the inconsistency. 
The transformation of the full $\pi$ field is given by
   $[  Q_5^{\rm LF}, \pi ]= 2i ( \sigma_0^{\rm (op)}+\varphi_\sigma )
    + \Delta \pi_0^{\rm (op)}\neq 2i\sigma $.
The vacuum expectation value of this equation gives a consistent result:
\begin{eqnarray}
\left\langle 0 \left| \left[  Q_5^{\rm LF},\ \pi \right] 
   \right| 0 \right\rangle  
&=& 2i\left\langle 0\left| \sigma_0^{\rm (op)}
  +\varphi_\sigma \right|0\right\rangle 
  +\left\langle 0\left|\Delta \pi_0^{\rm(op)}\right|0\right\rangle \nonumber\\
&=& 0.\label{bawr}
\end{eqnarray}
Now it is easy to obtain the transformation of $\bar\Psi i\gamma_5 \Psi$.
Our final result is 
\begin{eqnarray}
&&\left[ Q_5^{\rm LF} , \bar \Psi i \gamma_5 \Psi \right]
  = 2i \bar \Psi \Psi\\
&&\quad\quad    +\left\{ \psi_+^{a \dag} \frac{\gamma^-}{\sqrt{2}}
       \left( i\Delta\pi_0^{\rm (op)}+2  M \right)
      \frac{1}{2} \frac{1}{i \partial_-} \gamma^+ \psi^a_+
     +{\rm h.c.} \right\}.\nonumber
\end{eqnarray}
In addition to the first term that is equivalent to the usual result, 
  we have nonvanishing extra terms. 
However, if one takes the vacuum expectation value of this equation, 
  such extra terms should exactly cancel the first term 
  $2i\langle \bar\Psi\Psi\rangle\neq0$.
It is indeed the case and the explicit evaluation of the r.h.s. gives
  a consistent result 
\begin{equation}
 \left\langle 0 \left\vert \left[  Q_5^{\rm LF}, 
                                 \bar\Psi i\gamma_5 \Psi \right] 
 \right\vert 0 \right\rangle =0.
\end{equation}
Here we neglect the term $[ \Delta\pi_0^{\rm (op)}, \psi_+^{a} ] \sim O(1/N)$.
Thus we checked the consistency between the null-plane charge 
 $Q_5^{\rm LF}$ and the chiral condensate $\langle \sigma \rangle =
 -\frac{\lambda}{N}\langle \bar\Psi \Psi \rangle \neq 0$ upto 
 the mean-field level.

As the result of these unusual chiral transformation, 
  even the hamiltonian loses the chiral symmetry in the broken phase:
The commutator $\left[Q_5^{\rm LF}, H\right]\neq 0$ is directly evaluated 
  exactly in the same way as above.
This means that the LF chiral charge is not conserved in the broken phase.
It should be emphasized that the violation is proportional to 
  the fermion's dynamical mass $M$ and thus does not vanish 
  even in the chiral limit.
It is very interesting that the chiral symmetry breaking 
   in the LF formulation is expressed as an explicit breaking.
The important difference, however, is that usual explicit breaking 
  does not accompany the gap equation, while in our case the gap 
  equation plays a very important role in many aspects.
The nonconservation of $Q_5^{\rm LF}$ on the LF in the broken phase 
  has been discussed by several people in relation to PCAC 
  \cite{Carlitz,nonconserv,Tsujimaru-Yamawaki}.
Particularly, a similar situation to our conclusion (i.e. the 
  non-conservation of the null-plane charge in the DLCQ method) was 
  found in the broken phase of the scalar model \cite{Tsujimaru-Yamawaki}.
The problem of nonconserving charge should be intimately connected 
  with the divergence of the chiral current.
In Sec.~V-C, we will again meet the non-conservation of $Q_5^{\rm LF}$  
  as a result of the PCAC relation and peculiar behavior of the pion zero mode
  in the chiral limit.

Let us turn to the symmetric phase where the coupling is large 
  but slightly less than the critical value $\lambda \stackrel{<}{\sim} 
  \lambda_{\rm cr}$.
In this region, we use the symmetric solution of the gap equation. 
If we restrict ourselves to the chiral limit $m=0$, the solution is just 
  a trivial one $M=0$. 
Transformation law in this phase is obtained 
  by simply substituting $M=0$ into the above results.
Therefore in the leading order of $1/N$ expansion, all the dependent 
  fields transform in a chiral symmetric way:
\begin{eqnarray}
  & & \left[  Q_5^{\rm LF} , \sigma_0^{(\rm op)} \right]
        = -2 i \pi_0^{(\rm op)},     \ 
      \left[  Q_5^{\rm LF} , \pi_0^{(\rm op)} \right]
       =2 i  \sigma_0^{(\rm op)} ,      \\
  & & \left[ Q_5^{\rm LF} , \psi^a_- \right] 
     = \gamma_5 \psi^a_-.
  \label{kc}
\end{eqnarray}
With these commutators, we find also $\left[ Q_5^{\rm LF} , H \right]=0$.
Thus the LF chiral charge is conserved in the symmetric phase as expected.

\subsection{Masses of the scalar and pseudoscalar bosons}

Here we calculate the masses of scalars which are bound states of 
  fermion and antifermion.
Unlike the NJL model, we have 'dynamical' scalars in the lagrangian. 
Therefore it is convenient to evaluate the ``pole mass'' of the scalars 
    directly from their propagators 
   without considering bound-state equations.
The procedure of calculating the pole masses is as follows.
First we insert the broken or unbroken solutions $\sigma_0$, $\pi_0$, 
   and $\psi_-$ into the original lagrangian.
For simplicity, we ignore the finite $L$ effect.
This is because we are only interested in the effects of the condensate 
   $\sigma_0^{\rm (c)}$ and the nonzero constituent mass $M$.
Next, reading the fermion propagator from the lagrangian, we calculate the 
  scalar propagators $\Delta_{\pi,\sigma}(k^2)$ upto one loop of the 
  fermion.
Finally, the pole masses are obtained from the equation 
  $\Delta_{\sigma, \pi}^{-1}(k^2=m_{\sigma,\pi}^2)=0$.

\subsubsection{Broken phase}

Let us first consider the broken phase $\lambda>\lambda_{\rm cr}$.
Inserting the broken solutions into the lagrangian, 
  we have 
\begin{eqnarray}
  {\cal L}&=&{N \over 2 \mu^2} ( \partial_\mu \hat\sigma
            \partial^\mu \hat \sigma
           + \partial_\mu \hat \pi \partial^\mu \hat \pi )
          - {N \over 2 \lambda}\left\{ (\sigma_0^{(\rm c)} 
                              + \hat \sigma)^2+ \hat \pi^2\right\}\nonumber\\
   &&       + \sqrt2 \psi_+^{a \dag} i \partial_+ \psi^a_+   \nonumber\\
   &&     + \sqrt2 \Big\{  i \partial_- \psi_{-M}^{a \dag} 
       - {1 \over 2}  \psi_+^{a \dag} \gamma^- (\hat \sigma +i \hat \pi
                      \gamma_5)
                    \Big\}\nonumber\\
   &&  \hskip 1.5cm \times \Big\{  \psi^a_{-M}
        + {1 \over 2} \frac{1}{i \partial_-} 
           ( \hat \sigma -i \hat \pi \gamma_5 ) \gamma^+ \psi_+^a  \Big\} 
       \nonumber \\
   &=&   {N \over 2 \mu^2} ( \partial_\mu \hat\sigma
            \partial^\mu \hat \sigma
           + \partial_\mu \hat \pi \partial^\mu \hat \pi )
          - {N \over 2 \lambda} \left\{ (\sigma_0^{(\rm c)} + \hat \sigma)^2
            + \hat \pi^2\right\}                    \nonumber  \\
  & &  + \bar \Psi^a_M~ (~i \rlap/ \partial~ - M )\Psi^a_M~       
       - \bar \Psi^a_M~ ( \hat \sigma +i \hat \pi \gamma_5 )\Psi^a_M~     
                                            \nonumber  \\
  & & - {1 \over 2} \bar \Psi^a_M~ ( \hat \sigma +i \hat \pi
           \gamma_5 ) \gamma^+
        \frac{1}{i \partial_-} \left\{ ( \hat \sigma +i \hat \pi \gamma_5
)\Psi^a_M~ \right\}~,
  \label{jfc}
 \end{eqnarray}
where we used the notation 
   $\sigma(x)=\sigma_0^{(\rm c)}+\hat \sigma(x)$ and $\pi(x)=\hat\pi(x)$, 
   and $\Psi_M$ is defined by Eq.~(\ref{cec}).
Instead of the fermion propagator for + component  
$S_{++}(p)=\sqrt{2} \Lambda_+p_-/(p^2-M^2+i\epsilon)$,
it is convenient for practical calculation to define the  propagator 
   for $\Psi^a_M$: 
\begin{eqnarray}
   {\overline S}(p) &\equiv& \frac{ \overline {\rlap/ p} +M}
                       { p^2 -M^2 +i \epsilon }\  ,
  \label{jfd}
\end{eqnarray}
where ${\overline p}^{\mu}=( p^+, p^- = ( {\bf p}_{\perp}^2 +M^2 )/ 2 p^+,
    {\bf p}_{\perp})$ is the on-shell four momentum \cite{ChangRootYang}. 
Note that this partially on-shell propagator ${\overline S}(p)$ is 
  different from the usual fermion propagator 
  $S(p)$ by an instantaneous part 
  ${\overline S}(p)=S(p)-\frac{\gamma^+}{2 p^+}$, which arises from   
  the bad component $\psi_-$ as the solution of the fermionic constraint.  

Scalar and pseudoscalar propagators $\Delta_{\sigma, \pi}(k)$ 
  with fermion's one loop quantum correction are given by
\begin{eqnarray}
  \frac1N\left(\matrix{\Delta_\pi^{-1}(k)\cr \Delta_\sigma^{-1} (k)}
         \right)   
    = {k^2 \over \mu^2} -{1 \over \lambda}
         + F_{\rm Yukawa}(k) + F_{\rm inst}(k) ,
  \label{jfe}
\end{eqnarray}
where
\begin{equation}
   F_{\rm Yukawa}(k) =
    -  \int\left[d^4p\right]
           {\rm tr} \left[ \left(\matrix{i\gamma_5\cr 1}\right) 
                          {\overline S}(p) 
                           \left(\matrix{i\gamma_5\cr 1}\right) 
                          {\overline S}(p-k) \right] 
  \label{jff}
\end{equation}
comes from the Yukawa interaction (FIG.~2) and
\begin{eqnarray}
   F_{\rm inst}(k) =
    -\frac{1}{2} \int && \left[d^4p\right] 
     {\rm tr} \left[ {\overline S}(p) \left(\matrix{i\gamma_5\cr 1}\right)
                     \frac{\gamma^+}{p^+ -k^+} 
                              \left(\matrix{i\gamma_5\cr 1}\right)\right.
                           \nonumber\\
    +&&\left. {\overline S}(p) \left(\matrix{i\gamma_5\cr 1}\right)
                     \frac{\gamma^+}{p^+ +k^+}
			      \left(\matrix{i\gamma_5\cr 1}\right)
                         \right] 
\end{eqnarray}
from the instantaneous interaction (FIG.~3).
The integration measure is given by 
$$
\int\left[d^4p\right]= \int { d^2 p_{\bot} \over i (2 \pi)^2 }
            \int_{- \infty}^{\infty} { d p^{+} \over 2 \pi } 
          \int_{- \infty}^{\infty} { d p^{-} \over 2 \pi }.
$$
Summation over the longitudinal discrete momenta $p_n^+$ is approximated 
  by integration.

\begin{figure}[ht]
\begin{center}
\vspace*{1cm}
\psfig{file=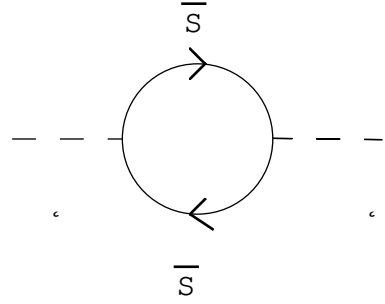}
\caption[]{Fermion's one loop contribution coming from the Yukawa interaction.
The solid line is for the fermion $\Psi_M$, the dashed line for $\pi$ or 
 $\sigma$.}
\end{center}
\end{figure}

\begin{figure}[ht]
\begin{center}
\vspace*{1cm}
\psfig{file=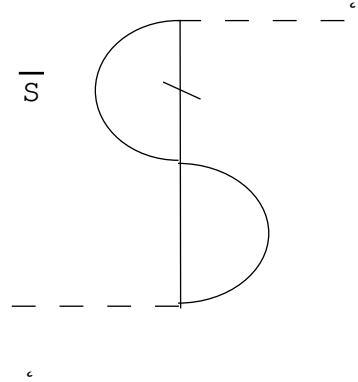}
\caption[]{Fermion's one loop contribution coming from the instantaneous 
 interaction which is represented by the vertical solid line.}
\end{center}
\end{figure}

For simplicity, we put $k_\bot =0 $.
Using a parameter $x\equiv p^+/k^+$, the propagators $\Delta_{\pi,\sigma}(k)$ 
 are expressed as
\begin{eqnarray}
&&\frac{1}{N}\left(\matrix{\Delta_\pi^{-1}(k)\cr \Delta_\sigma^{-1}(k)}\right) 
  = \frac{k^2}{\mu^2} - \frac{m^2_{\rm ZM}}{\mu^2}
          + {\cal F}(k^2, M^2) \left(\matrix{k^2\cr k^2-4M^2}\right) ,
    \nonumber\\
&&{\cal F}(k^2, M^2)\equiv\!\!  
      \int_{0}^{1}\!\! dx\! \int\! \frac{ d^2 p_{\bot} }{(2 \pi)^3 } 
           \frac{1}{p_{\bot}^2 +M^2-k^2 x(1-x)}
  \label{integral}
\end{eqnarray}
where we have utilized the gap equation  (\ref{cai}) and 
 $m_{\rm ZM}^2=\mu^2 m/\lambda M$.
The physical masses are determined from the equations 
   $\Delta_{\sigma,\pi}^{-1} (k^2=m_{\sigma, \pi}^2)=0$.
Since the integral in ${\cal F}(k^2, M^2)$ diverges, 
  we must specify a cutoff. 
Here we use the ``extended PI cutoff'' \cite{Bentz}:
\begin{equation}
\sum_i p_\pm^{(i)}<\Lambda,
\end{equation}
where $i$ denotes the particles of the internal lines.
This cutoff is a natural generalization of the naive PI cutoff 
  $p_\pm<\Lambda$ and can be applied to multiple internal lines. 
In our case, the extended PI cutoff becomes 
$$
\frac{p_\perp^2 + M^2}{x}+\frac{p_\perp^2 + M^2}{1-x}<2\Lambda^2.
$$
The explicit form of ${\cal F}(k^2, M^2)$ with this cutoff is given 
  in Appendix E.
Then we obtain 
  highly complicated nonlinear equations for $m^2_{\pi,\sigma}$ 
  which are also shown in Appendix E.
First of all, it is almost trivial that the equation for $m_\pi$ 
  has a solution $m_{\pi}=0$ in the chiral limit.  
For small bare mass $m \simeq 0$, we find 
\begin{equation}
   m_\pi^2 = \frac{m}{\lambda M_0} \left[ 
   \frac{1}{\mu^2} + {\cal F}_{\rm EPI}(k^2\rightarrow 0, M_0^2)
   \right]^{-1}  
  +O(m^2) ,
  \label{jk}
\end{equation}
where $M_0$ is a fermion condensate in the chiral limit\footnote{For 
small but finite $m$, the condensate is evaluated from Eq.~(\ref{caic}) 
as
$$
M=M_0+  \frac{\lambda_{\rm cr}}{\lambda} 
\frac{1}{\frac{M_0^2}{\Lambda^2}\ln \frac{2\Lambda^2}{M_0^2}}\ m
+ O(m^2).
$$
} and
  ${\cal F}_{\rm EPI}(k^2\rightarrow 0, M_0^2)$ is independent 
  of $m$ (see Appendix E).
Therefore we have checked that the mass of the pion goes to zero 
 in the chiral limit and now we can identify it with the NG boson.

The mass of $\sigma$ is determined in the same way. 
For example, in the chiral and heavy mass limit $m\rightarrow 0, 
\mu\rightarrow \infty$,  one can easily find a solution
 $m_\sigma^2=(2M)^2$ which is known to exist in the NJL model
 in the chiral limit.

Our result of the pion mass (\ref{jk}) satisfies the 
  Gell-Mann-Oakes-Renner (GOR) relation when $\mu\rightarrow \infty$:
\begin{equation}
 m^2_\pi f_\pi^2 = - 4 m \langle \bar \Psi \Psi \rangle,
\label{GOR}
\end{equation}
where $f_\pi$ is the pion decay constant. 
To see this, let us calculate $f_\pi$ explicitly.
Rewriting the pion's propagator in the chiral and heavy scalar limit as 
\begin{eqnarray}
  \Delta_\pi(k)&=&\frac{Z_\pi}{k^2-m_\pi^2},\label{propagator}\\
  Z_\pi &\equiv & 
  \left[N{\cal F}_{\rm EPI}(k^2\rightarrow 0, M_0^2)\right]^{-1},
\end{eqnarray}
we find the the effective pion-quark coupling $g_\pi=\sqrt{Z_\pi}$.
Then, in the leading order of $1/N$ expansion, $f_\pi$ is given by
 a fermion one-loop integral with appropriate Lorentz structure:
\begin{eqnarray}
ik^\mu f_\pi &\equiv & -\left\langle 0 \left\vert\  j^\mu_5(0)\ 
                 \right\vert \pi (k) \right\rangle\nonumber\\
             &=& g_\pi N \int \frac{d^4p}{(2\pi)^4}\ 
                 {\rm tr} \left[ \gamma^\mu \gamma_5
                          {\overline S}(p-k) \gamma_5 {\overline S}(p)
                          \right]\nonumber\\
	     &=&  2 i g_\pi N M_0 k^\mu 
                 {\cal F}_{\rm EPI}(k^2\rightarrow 0, M_0^2),
\label{f_pi_matrix}
\end{eqnarray}
therefore 
\begin{equation}
 f_\pi=2M_0\left[N{\cal F}_{\rm EPI}(k^2\rightarrow 0, M_0^2)\right]^{1/2}
      =\frac{2M_0}{\sqrt{Z_\pi}}\ .
 \label{f_pi}
\end{equation}
Using this result and $m_\pi^2=\frac{m}{\lambda M_0}NZ_\pi$, 
  we finally confirm the GOR relation (\ref{GOR}).

\subsubsection{Symmetric phase}

Next let us consider the symmetric case.
Using the symmetric solution to the zero-mode constraint, 
  we can evaluate the masses of 
 $\sigma$ and $\pi$ in the symmetric phase.
In the chiral limit, we have
\begin{equation}
 \frac{1}{N} {\Delta_{\pi,\sigma} (k)}^{-1}
 =  \frac{k^2}{\mu^2} -\frac{m^2_{\rm ZM}}{\mu^2}
   + {\cal F}(k^2, M^2=0) k^2
 \label{ke}
\end{equation}
where the zero-mode mass $m_{\rm ZM}^2$ in the symmetric phase is 
 expressed differently from that in the broken phase:
 $m_{\rm ZM}^2 = \frac{\mu^2}{\lambda}-\frac{\mu^2}{\lambda_{\rm cr}} + O(m)$.
Since $\Delta_{\sigma} (k)=\Delta_{\pi} (k)$, 
 $\sigma$ and $\pi$ have the same physical mass $m_\pi^2=m_\sigma^2$. 
The physical mass $m_\pi^2$ is obtained as a solution of 
\begin{equation}
   \frac{m_\pi^2}{\mu^2} +\left(\frac{1}{\lambda_{\rm cr}}
   -\frac{1}{\lambda} \right)
    = -m_\pi^2 \frac{\pi}{(2\pi)^3}\ln \left| 
         \frac{2\Lambda^2-m_\pi^2}{m_\pi^2} \right|.
  \label{kf}
\end{equation}
In the region $\lambda \stackrel{<}{\sim} \lambda_{\rm cr} $, 
  there is a nonzero solution $m_\pi^2 < 2\Lambda^2$.

In FIG.~4, we show the square masses of $\sigma$ and $\pi$ around 
   $\lambda\sim \lambda_{\rm cr}$ in the chiral and NJL limit 
   ($m\rightarrow 0, \ \mu\rightarrow \infty$).
In the broken phase ($\lambda> \lambda_{\rm cr}$), $m_\pi^2=0$ 
   and $m_\sigma^2=(2M)^2$, while in the symmetric phase 
   ($\lambda< \lambda_{\rm cr}$), $m_\sigma^2=m_\pi^2$ is given as 
   a solution to Eq.~(\ref{kf}).
The pion mass $m_\pi^2$ goes to zero in the limit
    $\lambda \rightarrow \lambda_{\rm cr}-0$.

\begin{figure}[hbt]
\begin{center}
\vspace*{1cm}
\psfig{file=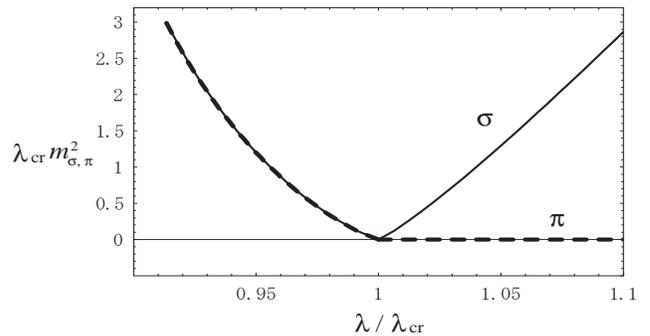, height= 5cm, width=8.5cm}
\caption[]{Squared masses $m^2_{\sigma, \pi}$ for $m\rightarrow 0$ 
  and $\mu\rightarrow \infty$  
  scaled by $\lambda_{\rm cr}^{-1}$.
Solid line is for $\sigma$ and dashed line for $\pi$.
For $\lambda > \lambda_{\rm cr}$  we used the broken solution, whereas 
for $\lambda < \lambda_{\rm cr}$   symmetric solution.}
\end{center}
\end{figure}

It is important to recognize that Eq.~(\ref{kf}) implies the existence 
  of tachyonic modes for $\lambda>\lambda_{\rm cr}$ as we mentioned before. 
Indeed, if we assume $\lambda>\lambda_{\rm cr}$, we find a negative 
  solution $m_\pi^2 < 0$.
Therefore, if we choose the symmetric solution to the zero-mode
  constraint, then the resulting theory becomes unstable for  
  $\lambda>\lambda_{\rm cr}$. 
So we must select the broken solution above the critical coupling.


\subsection{Derivation of the PCAC relation and the nonconservation of 
            $Q_5^{\rm LF}$}
As we discussed in Sec. III-B2, it is almost hopeless to treat the 
  LF chiral current $\widehat j_5^\mu$ for the massive case.
Instead, we adopt the current $j_5^\mu$ (\ref{btf}) which is much 
  more tractable than $\widehat j_5^\mu$ and gives the 
  same null-plane charge.  
Then it is straightforward to derive the PCAC relation. 
Consider the divergence of $j_5^\mu$ in the $\mu\rightarrow \infty$ limit
 [see Eq.~(\ref{divergence}) ]
\begin{eqnarray}
\partial_\mu j^\mu_5  
	&=& -2m\bar\Psi i \gamma_5 \Psi\nonumber\\
	&=& 2m \frac{N}{\lambda}\ \pi\nonumber\\
	&=& 2m \frac{N}{\lambda}\sqrt{Z_\pi}\ \pi_{\rm n},
\end{eqnarray}
where the normalized pion field $\pi_{\rm n}$ was introduced 
  so that its propagator be $\Delta_\pi^{\rm (n)}(k)=1/(k^2-m_\pi^2)$ 
  [see Eq.~(\ref{propagator})].
If we use the pion decay constant (\ref{f_pi}) and the pion mass (\ref{jk})
 [or equivalently $m_\pi^2=\frac{m}{\lambda M_0}NZ_\pi$], we obtain
 the PCAC relation,
\begin{equation}
\partial_\mu j^\mu_5 = m_\pi^2 f_\pi \pi_{\rm n}.
\label{PCAC}
\end{equation}
This is also consistent with Eq.~(\ref{f_pi_matrix}) [our normalization 
  is $\langle 0 | \pi_{\rm n}| \pi(k)\rangle =1$]. 

The important consequences of Sec.~V-A were that 
  i) the null-plane chiral charge $Q_5^{\rm LF}$ is not conserved 
  in the broken phase $\left[Q_5^{\rm LF}, H_{\rm LF}\right]\neq 0$, 
  and that ii) the violation is proportional to the dynamical fermion mass 
  $M$ and does not disappear even in the chiral limit.
Since we did not show explicitly the quantity because of its complexity,
  we here discuss it in more elegant way in the context of the PCAC relation.
First of all, let us remember that the chiral charge (\ref{btg}) 
  defined by $j_5^\mu$ is equivalent to the one defined from 
  $\widehat j_5^\mu$ due to $\widehat j_5^+=j_5^+$.
This means that the time derivative of $Q_5^{\rm LF}$ should be the same 
  even if we use the current $j_5^\mu$. 
Therefore, we can calculate the quantity $\partial_+ Q_5^{\rm LF}$
  from the PCAC relation (\ref{PCAC}):
\begin{eqnarray}
\partial_+ Q_5^{\rm LF}
	&=&\frac{1}{i}\left[Q_5^{\rm LF}, H_{\rm LF}\right]\nonumber\\
	&=&f_\pi m_\pi^2 \int dx^-d^2x_\perp  \pi_{\rm n}\nonumber \\
	&=&f_\pi m_\pi^2 \int dx^-d^2x_\perp  \pi_{\rm n}^0,
\end{eqnarray}
where $\pi_{\rm n}^0$ is the zero mode of $\pi_{\rm n}$.
If the current mass is not zero, the r.h.s. does not vanish in general 
  and the chiral charge $Q_5^{\rm LF}$ does not conserve.
Since the r.h.s. is proportional to $m_\pi^2$, it seems vanish in the 
  chiral limit $m_\pi^2 \propto m   \rightarrow 0$.
However, it does  survive finite  even in the chiral limit because 
 the pion zero mode shows the singular behavior in the chiral limit:
\begin{equation}
  \int dx^- d^2x_\perp \pi_{\rm n}^0 \propto \frac{1}{m}\ .\label{singular}
\end{equation}
Since $f_\pi$ is proportional to the fermion's dynamical mass 
  $M_0$ [Eq.~(\ref{f_pi})], we can confirm the result in Sec. V-A.
That is, the null-plane chiral charge $Q_5^{\rm LF}$ does not 
  conserve $\partial_+ Q_5^{\rm LF}\neq 0$  even in the chiral limit. 
The singular behavior of Eq.~(\ref{singular}) in the chiral limit
  has been pointed out by Tsujimaru and Yamawaki for spontaneously 
  broken theories\cite{Tsujimaru-Yamawaki}. 
They showed the necessity of introducing the nonzero mass 
  $m_{\rm NG}\neq 0$ for the NG boson in the broken phase 
  and found the singular behavior of the NG-boson zero mode 
  $\phi_0^{\rm NG}$ as
$$
\int dx^-d^2x_\perp \phi_0^{\rm NG} \sim \frac{1}{m_{\rm NG}^2}.
$$
We have confirmed their result for the $dynamically$ broken theory.

Now let us verify the singular behavior (\ref{singular}).
It is generally known that the zero-mode constraint for 
   the NG boson becomes inconsistent in the broken phase 
   unless we introduce finite mass of the NG boson by hand 
   as regularization \cite{Tsujimaru-Yamawaki}.
Using the zero-mode mass (\ref{ZM_mass}),
   the zero-mode constraint for $\pi_0$ is simply written as 
\begin{equation}
\left(m_{\rm ZM}^2-\partial^2_\perp\right)\pi_0
   =\int_{-L}^L dx^- \mu^2 f(x^-, x_\perp).
\end{equation}
Suppose $m_{\rm ZM}=0$ and introduce the periodic boundary condition on 
   $\pi_0$ in the transverse directions, then the transverse integral 
   $\int d^2x_\perp$ of the zero-mode constraint leads to 
   inconsistency\footnote{The mean-field result avoids this inconsistency 
   due to $\int d^3x\bar \Psi_M i\gamma_5\Psi_M=0$. However, 
   higher order calculation requires nonzero bare mass.} 
   $0=\int d^2x_\perp \int_{-L}^L dx^- \mu^2 f(x^-, x_\perp)\neq 0$, 
   which suggests to introduce ``zero-mode mass'' $m_{\rm ZM}\neq 0$.
In our calculation, the origin of the the finite mass of the NG boson 
   is the fermion's bare mass $m$ [not the scalar mass $\mu/\sqrt{\lambda}$]. 
Indeed, $m_{\rm ZM}^2=\mu^2m/\lambda M$ in the broken phase.
Therefore in the $\mu\rightarrow \infty$ limit, we find the singular
   behavior of the zero mode:
\begin{eqnarray}
\pi_0&=&\frac{1}{m_{\rm ZM}^2-\partial_\perp^2}
        \int_{-L}^L dx^- \mu^2 f(x^-, x_\perp)\nonumber\\
&\rightarrow& \frac{\lambda M}{m}\int_{-L}^L dx^- f(x^-, x_\perp).
\end{eqnarray}
On the other hand, in the symmetric phase $m_{\rm ZM}^2=
   \mu^2(\frac{1}{\lambda}-\frac{1}{\lambda_{\rm cr}})+O(m)$ 
   survives finite in the chiral limit.
We called $m_{\rm ZM}$ ``zero-mode mass'', but it should not be 
   confused with the physical pion mass $m_\pi$.
Both become nonzero due to nonzero bare mass $m\neq 0$, 
   but we have to calculate fermion's one loop 
   to obtain the physical pion mass $m_\pi$.

\setcounter{equation}{0}

\section{Conclusion and discussions}
We have studied a method of describing the dynamical 
  chiral symmetry breaking on the LF.
Our description is based on the idea in DLCQ that 
  the symmetry breaking is achieved by solving the ``zero-mode 
  constraints'', which already succeeded to some extent in describing the 
  spontaneous symmetry breaking in simple scalar models.
The point is that we can utilize this idea even for the dynamical 
  symmetry breaking 
  in fermionic systems if we introduce bosonic auxiliary fields 
  for $\bar\Psi\Psi$ and $\bar\Psi i\gamma_5\Psi$, and treat them as
  dynamical variables by adding their kinetic terms.
Then the problem can be formulated such that we find a nontrivial 
  solution to the  ``zero-mode constraint''.  
We exemplified this idea in the NJL model.
The model we studied is the chiral Yukawa model, 
  which reproduces the NJL model in the infinitely heavy mass limit of 
  the scalars.
Within this model, we showed in the massless case the equivalence 
  between ``light-front'' chiral transformation and the usual one 
  by classically solving the three (i.e. two zero-mode and one fermionic) 
  constraints.
This allowed us to construct the chiral current $j^5_{\mu}$ and 
  charge $Q_5^{\rm LF}$. 
Even if we solve the constraints classically, the resulting theory 
  cannot have a symmetry breaking term. 
Quantum analysis showed  
  that the zero-mode constraint for a scalar boson $\sigma$ 
  became the gap equation in nonperturbative treatment, which  
  lead to nonzero condensate $\langle \sigma \rangle \neq 0$ 
  and equivalently to the chiral symmetry breakdown. 
We found the critical coupling $\lambda_{\rm cr}$ beyond 
  which the fermion acquires nonzero dynamical mass.
On the other hand, a perturbative solution could not 
  give fermion condensate even in the quantum theory.

The most important key of our description was the identification of 
  the zero-mode constraint of $\sigma$ with the gap equation.
This was suffered from a severe problem that the correct mass dependence 
  disappears from the mode expansion.
Of course this is a demerit of the LF formalism and we have to 
  carefully incorporate mass dependence into e.g. 
  $\langle \bar\Psi\Psi \rangle$ when we regularize its infrared divergence.
It is suggestive that cutoff schemes with symmetry consideration 
  such as parity or rotational invariance lead to 
  physically acceptable results.
Contrary to such negative aspects, the mass-information loss has 
  a useful and important aspect. 
It follows that the Fock vacuum keeps invariant even if we change 
  the value of mass.  
Therefore the vacuum does not change even though symmetry breaking 
 occurs and the fermion acquires dynamical mass.

In our formalism, {\it the vacuum is exactly the Fock vacuum}.
The inclusion of {\it dynamical} scalar fields was necessary to clarify  
  the structure of the Hilbert space and the triviality of the vacuum.
The way of realizing broken phase is that 
  the vacuum is still trivial but the operator 
  structure of the {\it dependent} variables changes.
In other words, the ``vacuum physics'' in conventional formulation
  is converted into the hamiltonian through the dependent variables. 
The zero modes of scalars and the bad component $\psi_-$ are constrained 
  variables and differently expressed by physical variables depending 
  on the phases.
Related to this, the LF chiral transformation of the dependent 
  variables also becomes unusual in the broken phase.  
Consequently, a seemingly contradiction between the triviality 
  of the null-plane charge $Q_5^{\rm LF}|0\rangle =0$ and 
  the chiral condensate $\langle \bar \Psi \Psi \rangle\neq 0$ 
  is resolved.

We further calculated  masses of $\pi$ and $\sigma$ for 
  both symmetric and broken phases. 
In the broken phase, the mass of $\pi$ goes to zero in the chiral limit, 
  which is consistent with the NG theorem.
Our result is consistent with the GOR relation. 
If we substitute symmetric solution into the lagrangian, there appear
  tachyon modes for $\lambda>\lambda_{\rm cr}$.
Therefore we can say that when  $\lambda>\lambda_{\rm cr}$, 
  physically realized phase is the broken phase. 
Certainly we have massless pion in the model, but it is very difficult 
  to verify the NG theorem in general on the LF.
This is because we have nonlocal interaction and because 
  the chiral transformation of the full fields explicitly 
  depends on the model.
Both of these arise from the fact that in the LF formalism 
  the bad component of fermion and zero modes of scalars 
  are constrained variables. 
However, we have succeeded in deriving the PCAC relation.
This was enabled by utilizing the current $j_5^\mu$ which was 
  the Noether current defined for the massless fermion.
The 'massive' current $\widehat j_5^\mu$ becomes complicated and 
  it is almost hopeless to deal with it.
The physics meaning of this discrepancy between $j_5^\mu$ and 
  $\widehat j_5^\mu$ is still not clear but it seems that 
  the usual current $j_5^\mu$ is favorable for discussing the 
  ``usual'' chiral symmetry (not the ``LF'' chiral symmetry).

One of the most important conclusion of our analysis is the 
  nonconservation of the LF chiral charge $Q_5^{\rm LF}$.
This can be shown both by direct calculation of 
  $[Q_5^{\rm LF}, H_{\rm LF}]$ using the unusual chiral transformation law
  and  by utilizing the PCAC relation.
In the broken phase, the chiral charge $Q_5^{\rm LF}$ 
  does not conserve $\partial_+ Q_5^{\rm LF}\neq 0$ 
  even in the chiral limit.
The singular behavior $1/m$ of the pion zero mode  is essential to 
  give a finite violation of $\partial_+ Q_5^{\rm LF}\neq 0$.

In our calculation, ``nonperturbative'' implied the mean-field approximation. 
This mean-field calculation is justified  as the leading 
  order approximation in the  $1/N$ expansion.
In principle, we can develop a systematic $1/N$ expansion to go 
  beyond the mean-field result.
Nevertheless, the higher order will severely depend on the operator ordering 
  and it is not clear whether the result with our specific ordering 
  makes sense.
If we want to go beyond the leading order, we have to determine the 
  consistent operator ordering according to the criterion discussed 
  in the text. 
Since this is a very difficult task in our model, it should be 
  examined in much simpler models such as 1+1 dimensional Yukawa theory.

It will be challenging to use other nonperturbative method to solve 
  the zero-mode constraint.
For example, Tamm-Dancoff approximation which truncates the Fock space 
into a few particle states will give some nontrivial results.
Notice that our leading $1/N$ approximation corresponds to 
  the two body truncation since multiquark states give higher order 
  contribution.

Our method here  heavily relies on the introduction of auxiliary fields.
So it seems natural to ask a question:
Can we describe the chiral symmetry breaking 
 without introducing the auxiliary fields?
The answer is of course yes.
Even though we do not have zero mode constraints, it is possible to 
  describe the chiral symmetry breaking on the LF.
Its explicit demonstration in the purely fermionic NJL model will be given 
 in our next paper \cite{II}. 
As far as the NJL model is concerned, to solve the fermionic constraint 
 becomes of great importance.

There still remain many problems which cannot be discussed in our model.
One of them is the issue of renormalization.
In a renormalizable theory, if one introduces an infrared cutoff 
  and excludes the zero mode degrees of freedom from the beginning, 
  then the ``vacuum physics'' should be discussed as the problem of 
  renormalization with nonperturbative infrared counter terms.
Relation between such counter term approach \cite{Burkardt} and 
  the zero-mode approach presented here is not clear. 
We need further investigation for understanding how to describe the 
  chiral symmetry breaking in LFQCD.

\section*{Acknowledgments}

The authors acknowledge W. Bentz for discussions on the  
  cutoff scheme. 
One of them (K.I.) is thankful to K. Yazaki and K. Yamawaki 
  for useful discussions and to members of Yukawa Institute 
  for Theoretical Physics where most of the work was done.
The other author (S.M.) is grateful to the members of 
  Saturday Meeting (Doyo-kai) for stimulating discussions.


\appendix

\setcounter{equation}{0}
\section{The chiral Yukawa model}

Effective potential for scalars in the chiral Yukawa model (\ref{aa}) 
  is easily calculated in the leading order of $1/N$ expansion \cite{Kugo}.
Exactly in the same way as in the Gross-Neveu model \cite{GN}, 
  we find that the leading contribution comes from the fermion 
  one-loop diagrams.
Note that the inclusion of kinetic terms for scalars has no effect on the 
  leading effective potential.  
Since the scalar propagator is $O(1/N)$, effects of the kinetic term 
  (i.e. $\mu$ dependence) emerges from the next leading order.
The effective potential $V(\sigma, \pi)$ in the leading order 
  is independent of $\mu$ and is given by
\begin{eqnarray}
V(\sigma, \pi)/N = &&  \frac{1}{2\lambda}(\sigma^2+\pi^2) \nonumber\\
                && +\ i \int \frac{d^4k}{(2\pi)^4} 
                     \ln \det [k\!\!\!/ - (\sigma + i\gamma_5\pi)],
\end{eqnarray}
which is the same result as that of the NJL model.
Therefore evaluating the integral and differentiating $V(\sigma, \pi=0)$ 
 with respect to $\sigma$, we obtain the gap equation which determines 
 the vacuum expectation value of $\sigma$.

Now let us turn to the mean-field approximation. 
Using $AB\approx \langle A \rangle B + A \langle B \rangle 
- \langle A \rangle\langle B\rangle $, the Yukawa interaction becomes 
\begin{eqnarray}
&&-{\cal L}_{\rm Yukawa}\nonumber\\
&&\quad \approx \langle \sigma \rangle \bar \Psi \Psi 
                               + \sigma \langle \bar \Psi \Psi \rangle
                        + \langle \pi \rangle \bar \Psi i \gamma_5 \Psi 
                        + \pi \langle \bar \Psi i\gamma_5 \Psi \rangle
			+\  {\rm const.},\nonumber
\end{eqnarray}
where $\langle \sigma \rangle =O(N^0), \ \sigma - \langle \sigma 
\rangle =O(N^{-1/2})$ and similarly for others.
The leading order  Euler-Lagrange equations 
  for $\sigma$ and $\pi$ are
  $\langle \sigma \rangle = -\frac{\lambda}{N}\langle \bar \Psi \Psi \rangle$  
  and $\langle \pi \rangle
       =-\frac{\lambda}{N}\langle \bar \Psi i\gamma_5 \Psi \rangle$.
Therefore, in the mean-field approximation 
 (=leading order of $1/N$ expansion), the chiral Yukawa model allows 
 both fermion and scalar condensates.
However, in higher order, fermion bilinears and scalars can independently 
 take their VEVs and the same relations do not necessarily hold.
The Euler-Lagrange equation for fermion in the mean-field approximation
  becomes $(i\partial\!\!\!/ -m -\langle \sigma \rangle - 
  \langle \pi \rangle i \gamma_5)\Psi=0$. 
Evaluating the fermion condensate in a self-consistent way, we obtain 
  the gap equation.

\setcounter{equation}{0}
\section{Conventions}

We summarize our convention. 
We follow the Kogut-Soper convention \cite{Review}.
First of all, the light-front coordinates are defined as
\begin{equation}
x^\pm = \frac{1}{\sqrt2}(x^0\pm x^3), \quad
x^i_\perp = x^i \quad(i=1,2), 
\end{equation}
where we treat $x^+$ as ``time''. 
The spatial coordinates $x^-$ and $x_\perp$ 
  are called the longitudinal and transverse 
  directions respectively. 
Derivatives in terms of $x^\pm$ are defined by
\begin{equation}
\partial_\pm\equiv \frac{\partial}{\partial x^\pm}.
\end{equation} 
For the $\gamma$ matrices, we also define  
\begin{equation}
\gamma^\pm=\frac{1}{\sqrt2}(\gamma^0\pm\gamma^3).
\end{equation}
It is useful to introduce projection operators $\Lambda_\pm$ 
  defined by
\begin{equation}
\Lambda_{\pm}=\frac12 \gamma^\mp \gamma^\pm
=\frac{1}{\sqrt2}\gamma^0\gamma^\pm.
\end{equation}
Indeed $\Lambda_\pm$ satisfy the projection properties 
$\Lambda_\pm^2=\Lambda_\pm,\ \Lambda_++\Lambda_-=1$, etc.
Splitting the fermion by the projectors,
\begin{equation}
  \Psi^a = \psi^a_+ + \psi^a_- , \quad 
  \psi^a_{\pm} \equiv \Lambda_{\pm}\Psi^a ,
\end{equation}
we find that for any fermion on the LF, 
  $\psi_-$ component is a dependent degree of freedom. 
$\psi_+$ and $\psi_-$ are called the ``good component'' and 
  the ``bad component'', respectively.

In DLCQ,  we set $x^-$ finite $x^-\in [-L, L]$ with some 
  boundary conditions on fields.
Taking the periodic boundary condition, we can clearly separate 
  a longitudinal zero mode from oscillating modes.
The zero mode of some local function $f(x)$ is defined by
\begin{equation}
f_0(x_\perp)=\frac{1}{2L}\int_{-L}^L dx^- f(x^-,x_\perp).
\end{equation}
The rest is the oscillating part: 
\begin{equation}
\varphi_f(x)=f(x)-f_0(x_\perp).
\end{equation}
For some composite fields, we use the notation 
  $[\quad]_0$ for their zero modes:
\begin{equation}
\Big[f(x)g(x)\Big]_0=\frac{1}{2L}\int_{-L}^L dx^- f(x)g(x).
\end{equation}
The inverse of the differential operator $\partial_-$ is defined as
\begin{equation}
\frac{1}{\partial_-}f(x^-) \equiv \int_{-L}^L dy^- \frac12\epsilon(x^--y^-)
f(y^-),
\end{equation}
where $\epsilon(x^-)$ is a sign function
\begin{equation}
\epsilon(x^-)=\left\{ \matrix{1 \quad (x^->0)\cr 
                              0 \quad (x^-=0)\cr 
                             -1 \quad (x^-<0)}
\right. .
\end{equation}

\setcounter{equation}{0}
\section{Constraints and Their Classical Solutions}

Since systems on the LF always have several constraints, 
 the LF quantization must be performed using Dirac's Hamiltonian 
 formalism for constrained systems.

The chiral Yukawa model has six primary constraints.
Among them, consistency conditions for $\theta_1=\Pi_{\sigma_0}\approx 0$ 
  and $\theta_2=\Pi_{\pi_0}\approx 0$ 
  ($\Pi_\Phi$ is a conjugate momenta of $\Phi$) generate the zero-mode 
  constraints (\ref{acc}), while $\theta_3=\Pi_{\psi_-}\approx 0$ 
  generates the  fermionic constraint (\ref{acaa}).
The consistency is calculated by $\dot \theta_i = \{\theta_i, 
  H_{\rm LF}+\lambda_j\theta_j \}_{\rm PB}\approx 0 $ 
  with the canonical light-front Hamiltonian $H_{\rm LF}=P^-$
\begin{eqnarray}
 &P^-&  = P^-_{\rm F}+P^-_{\rm B}+P^-_{\rm Y},   \label{ham} \\
 &&\hspace*{-0.5cm}P^-_{\rm F}= \int d^3x  
        \left\{ -  \bar \Psi^a i \gamma^-  \partial_- \Psi^a  
                - \bar \Psi^a i \gamma^\perp \partial_\perp\Psi^a
                + m \bar \Psi^a \Psi^a  
        \right\}       ,    \nonumber                     \\
 &&\hspace*{-0.5cm}P^-_{\rm B}= \int d^3x \biggl[ - {N \over 2 \mu^2} 
        \left\{  ( \partial^\perp\sigma_0\partial_\perp\sigma_0
                +  \partial^\perp\varphi_\sigma\partial_\perp\varphi_\sigma
                     )\right.\biggr. \nonumber\\
  &&    \hskip 2.8cm  \left. + ( \partial^\perp \pi_0 \partial_\perp \pi_0
                  + \partial^\perp\varphi_\pi\partial_\perp\varphi_\pi 
                     ) 
       \right\}                           \nonumber \\
   & & \hskip 1.9cm \biggl. + {N \over 2 \lambda} 
        \left\{ ( \sigma_0^2 + \varphi_\sigma^2 )
              + ( \pi_0^2 + \varphi_\pi^2 ) 
        \right\}    \biggr]     ,     \nonumber            \\
 &&\hspace*{-0.5cm}P^-_{\rm Y}=  \int d^3x
              \left\{ ( \sigma_0 + \varphi_\sigma ) \bar \Psi^a \Psi^a
              + ( \pi_0 + \varphi_\pi ) \bar \Psi^a i \gamma_5 \Psi^a
              \right\}       , 
      \nonumber 
\end{eqnarray}
where $\int d^3x=\int_{-L}^Ldx^- \int d^2x_\perp$.
As a result, we find that no more constraints are generated 
 from consistency conditions and that 
 this system belongs to the second class.

If we ignore ordering of the variables, it is not difficult to 
   solve the constraints. 
First, the fermionic constraint (\ref{acaa})  is solved as 
\begin{equation}
    \psi^a_- = \frac{1}{i \partial_-} \left[{1 \over 2} 
           \left(i\gamma^\perp\partial_\perp + m 
                 + \sigma - i\pi\gamma_5
           \right) 
           \gamma^+ \psi^a_+\right] ~,
    \label{acaab}
\end{equation}
where $\partial_-^{-1}$ is defined so that $\psi_-(x)$ also satisfies 
   the antiperiodic boundary condition (see Appendix B).
Note that however this solution still contains the zero modes 
   $\sigma_0$ and $\pi_0$ and thus is not a complete solution. 
Substituting (\ref{acaab}) into  Eq.~(\ref{acc}), 
   we have equations only for $\sigma_0$ and $\pi_0$. 
Then the formal solution is given by
\begin{eqnarray}
    && \left(
	\matrix{\sigma_0 (x_\bot) \cr \pi_0 (x_\bot)}
       \right)\nonumber\\
    &&  = -{1 \over \sqrt{2}}{\mu^2 \over N} {\cal D}^{-1}(x_\perp)
	\left[
	     \psi_+^{a \dag}\left(\matrix{-1\cr i\gamma_5}\right)
		\frac{1}{i \partial_-}(i\gamma^\perp \partial_\perp-m)
               \psi^a_+
	\right.					\nonumber\\
    && \quad\quad \left.  +\ \psi_+^{a \dag} \frac{1}{i \partial_-}
		\left\{ \left(\matrix{
			   \varphi_{\sigma}+i\gamma_5\varphi_\pi \cr
			   \varphi_{\pi}-i\gamma_5\varphi_\sigma}	
			\right)\psi_+^a
		\right\}+\ {\rm c.c.}\ 
 	\right]_0 ,
  \label{btb}
\end{eqnarray}
where the transverse differential operator ${\cal D}(x_\perp)$ is 
$$
  {\cal D}(x_\perp)
          ={\mu^2 \over \lambda}
          +{\mu^2 \over N}{1 \over \sqrt{2}}
             \left[ \psi_+^{a \dag} \frac{1}{i \partial_-} \psi^a_+
                    -\left(\frac{1}{i \partial_-}\psi_+^{a \dag}\right) 
                      \psi^a_+
             \right]_0
          -\partial_\bot^2 .
$$
The final expression for $\psi_-$ is reached after we insert
  Eq.~(\ref{btb}) into Eq.~(\ref{acaab}).
Though we have completely ignored the ``ordering'' in the classical treatment,
  the operator ordering becomes an issue in a quantum theory, which 
  makes the analysis very complicated.

\setcounter{equation}{0}
\section{Perturbative Solution to the Zero-Mode Constraints 
         in Quantum Analysis}

Let us solve the zero-mode constraints in perturbation theory 
   with the natural ordering in Eqs.~(\ref{acaa}) and (\ref{acc}). 
In addition to the expansions (\ref{exp1})-(\ref{exp3}),
  it is convenient to define the expansion of $\Psi$:
\begin{equation}
 \Psi^a = \sum_{n=0}^{\infty}
             \left(\frac{\lambda}{\lambda_{\rm cr}}\right)^n\Psi^{a (n)},  
  \label{bb} 
\end{equation}
where $\Psi^{a(0)} = \psi^a_+ + \psi_-^{a (0)}$ and 
 $\Psi^{a(k)} = \psi_-^{a (k)}$ for $k \ge 1$.

Knowing the lowest order solutions (\ref{bab}), we obtain 
  the next order solution,
\begin{eqnarray}
&&  \left(\matrix{\sigma_0^{(1)} \cr \pi_0^{(1)} } \right)
  =  -{\lambda_{\rm cr} \over N} \left[
         \bar \Psi^{a (0)}\left(\matrix{1\cr i\gamma_5}\right) 
              \Psi^{a (0)}   
   \right]_0 ~,
  \label{bbc}\\
&&    \psi_-^{a (1)}={1 \over 2}{1 \over i \partial_-}
    \left( \sigma_0^{(1)} - i \pi_0^{(1)} \gamma_5 \right) \gamma^+ \psi^a_+ .
  \label{bbe}
\end{eqnarray}
Similarly, if we know the solution up to $n$-th order, we easily obtain the 
  $(n+1)$-th order solution because the constraint equation is 
  written as follows
\begin{eqnarray}
 \frac{\mu^2}{\lambda_{\rm cr}} 
  \left(\matrix{ \sigma_0^{(n+1)} \cr \pi_0^{(n+1)} }\right)
  & =&  \partial_\perp^2 
  \left(\matrix{ \sigma_0^{(n)} \cr \pi_0^{(n)} }\right)\nonumber\\
  &- &{\mu^2 \over N} 
   \left[
     \bar \Psi^{a (0)}
       \left(\matrix{1\cr i\gamma_5}\right) 
     \Psi^{a (n)}   + {\rm h.c.}  
   \right]_0 ,
  \label{bcd}
\end{eqnarray}
where $\Psi^{a (n)}$ for $n\neq 0$ is 
\begin{equation}
 \Psi^{a (n)}=\frac12 \frac{1}{i\partial_-}
 \left(\sigma_0^{(n)}-i\pi_0^{(n)}\gamma_5\right)\gamma^+\psi_+^a,
 \label{bcdd}
\end{equation}
and we have used $\psi_-^{(k)\dag}\gamma^0\psi_-^{(l)}=0$ 
   for $k, l\neq 0$, etc.
In this way, we can determine the solution order by order.

\setcounter{equation}{0}
\section{Nonlinear Equations for Pole Masses}

With the Extended Parity Invariant (EPI) cutoff, the integral ${\cal 
F}(k^2,M^2)$  Eq.~(\ref{integral})  is given as
\begin{eqnarray}
  {\cal F}_{\rm EPI}(k^2, M^2)&\equiv&  
  \int_{x_{(-)}}^{x_{(+)}} dx 
  \int_0^{2\Lambda^2 x(1-x)-M^2} d p_{\bot}^2 \nonumber\\  
 &&  \quad   \times \  \frac{ \pi }{(2 \pi)^3 }
   \frac{1}{p_{\bot}^2 +M^2-k^2 x(1-x)}.
\end{eqnarray}
where integration limit 
  $x_{(\pm)}=(1\pm \beta )/2$ with  $\beta=\sqrt{1-2M^2/\Lambda^2}$
  comes from  $2\Lambda^2 x(1-x)-M^2>0$.
The integral is easily performed and the result is
\begin{eqnarray}
  &&{\cal F}_{\rm EPI}(k^2, M^2) = \frac{1}{8\pi^2}\log 
  \left( \frac{1+\beta}{1-\beta}\right)\nonumber\\
  &&\quad -\frac{1}{4\pi^2}\sqrt{\frac{4M^2-k^2}{k^2}}\arctan 
  \beta \sqrt{\frac{k^2}{4M^2-k^2}} .
\end{eqnarray}
For $k^2\sim 0$, we can approximate this as
\begin{equation}
  {\cal F}_{\rm EPI}(k^2\rightarrow 0, M^2)= \frac{1}{8\pi^2}\log 
  \left( \frac{1+\beta_0}{1-\beta_0}\right)
  -\frac{1}{4\pi^2} \beta_0,
\end{equation}
where $\beta_0=\sqrt{1-2M_0^2/\Lambda^2}$ and 
  we used $\lim_{x\rightarrow 0} x^{-1}\arctan x  =1$.

Using above, the nonlinear equations for the scalar masses are
\begin{equation}
\frac{m}{\lambda M}
      =\frac{1}{\mu^2}\left(\matrix{m_\pi^2\cr m_\sigma^2}\right)
       +\left(\matrix{m_\pi^2 {\cal F}_{\rm EPI}(m_\pi^2, M^2)\cr 
               (m_\sigma^2 -4M^2){\cal F}_{\rm EPI}(m_\sigma^2, M^2)}
        \right).
\label{nonlinear}
\end{equation}
Note that when the chiral and heavy mass limit $m\rightarrow 0$, $\mu 
\rightarrow \infty$, we have solutions $m_\pi =0$ and $m_\sigma = 2M$.


\end{document}